\documentclass[twocolumn]{aastex62}

\graphicspath{{./}{figures/}}

\received{TBD}
\revised{TBD}
\accepted{TBD}
\submitjournal{ApJ}

\shorttitle{Fundamental Properties of M dwarfs}
\shortauthors{Iyer et al}

\begin{document}


\title{The {\tt SPHINX} M-dwarf Spectral Grid. I. Benchmarking New Model Atmospheres to Derive Fundamental M-Dwarf Properties}

\correspondingauthor{Aishwarya Iyer}
\email{aiyer13@asu.edu}

\author[0000-0003-0971-1709]{Aishwarya R. Iyer}
\affil{School of Earth and Space Exploration,
Arizona State University,
525 E. University Dr., Tempe AZ 85281}
\affil{NASA FINESST Fellow}

\author[0000-0002-2338-476X]{Michael R. Line}
\affil{School of Earth and Space Exploration,
Arizona State University,
525 E. University Dr., Tempe AZ 85281}

\author[0000-0002-0638-8822]{Philip S. Muirhead}
\affil{Department of Astronomy and Institute for Astrophysical Research,
Boston University,
725 Commonwealth Ave., Boston, Massachusetts, 02215}

\author[0000-0002-9843-4354]{Jonathan J. Fortney}
\affil{Department of Astronomy and Astrophysics,
University of California, Santa Cruz,
1156 High St, Santa Cruz, CA 95064}

\author[0000-0002-4088-7262]{Ehsan Gharib-Nezhad}
\affil{NASA Ames Research Center, Moffet Field, CA. 94035 USA}
\affil{Bay Area Environmental Research Institute, CA, USA}



\begin{abstract}
About 70-80$\%$ of stars in our solar and galactic neighborhood are M dwarfs. They span a range of low masses (0.08 - 0.6 solar mass) and low temperatures (between 2500 K - 4000 K), facilitating molecule formation throughout their atmospheres. Standard stellar atmosphere models primarily designed for FGK stars have challenges in accurately characterizing broadband molecular features in spectra of cool stars. Here, we introduce $\tt{SPHINX}$---a new low-resolution (R$\sim$250) 1-D self-consistent radiative-convective thermochemical equilibrium chemistry grid of model atmospheres and spectra for characterizing M dwarfs. We incorporate the most up-to-date pre-computed absorption cross-section data with appropriate pressure-broadening treatment for key molecules dominant in late-K, early/main-sequence-M star atmospheres in our grid. We then validate our grid models by acquiring fundamental properties (T$_{eff}$, log$g$, metallicity, radius and C/O ratio) for a sample of 20 stars---10 benchmark M+G binary stars with known host star metallicities and 10 M dwarfs with interferometrically measured angular diameters. Incorporating a Gaussian-process inference tool $Starfish$, we also account for correlated and systematic noise sources in our grid-model fitting routine for low-resolution (R$\sim$120) (spectral stitching of SpeX, $SNIFS$, and HST $STIS$) observations and derive robust estimates of fundamental M dwarf atmospheric parameters. We also assess the influence of stellar photospheric heterogeneity on the acquired metallicity for a couple targets and find that could explain some deviations from observations. Moreover, we also probe whether the model-assumed mixing-length parameter for convection influences inferred radii, stellar effective temperature, and [M/H] to explain possible discrepancies compared to interferometry data and again find that this may bridge the gap between observations and model-derived stellar parameter deviations for cooler M dwarfs. Mainly, we show the unique strength in leveraging broadband molecular absorption features occurring in low-resolution M dwarf spectra and demonstrate the ability to improve constraints on fundamental stellar properties of exoplanet hosts and late brown dwarf companions.

\end{abstract}


\section{INTRODUCTION} \label{sec:intro}
M dwarfs are the most ubiquitous stars in our galaxy, solar neighborhood, and the universe put together, making up a majority of the combined stellar population by number \citep{Henry1994}. They have incredibly long main-sequence lifetimes, longer than the age of the universe, which make them extremely valuable for understanding the dynamical and chemical evolution of the galaxy (\citealt{bochanski2010,bochanski2011,lepine2003a, lepine2003b, lepine2013,lepine2011,west2011, hejazi2015}). Numerous space and ground-based exoplanet finding surveys have discovered many terrestrial planets orbiting M dwarf host stars \citep{gillon2017seven,nutzman2008design,irwin2009gj}. Currently, the Transiting Exoplanet  Survey  Satellite  (TESS)  is ongoing for the discovery of thousands of terrestrial exoplanets, most of them orbiting  M dwarfs  (\citealt{muirhead2018,ballard2019})—with the occurrence rate of at least $\sim$1 rocky planet around such a host \citep{dressing2013,dressing2015occurrence}. Because of a high occurrence of earth-like terrestrial planets in their habitable zones (\citealt{dressing2013,kopparapu2013,petigura2013,mulders2015,gaidos2016}), M dwarfs are valuable targets in the $JWST$ Era for understanding the hosting environment, formation, evolution and fundamental atmospheric properties of companion planets. 

\begin{figure}[!tbp]
    \includegraphics[width=\columnwidth, height=\textheight, keepaspectratio]{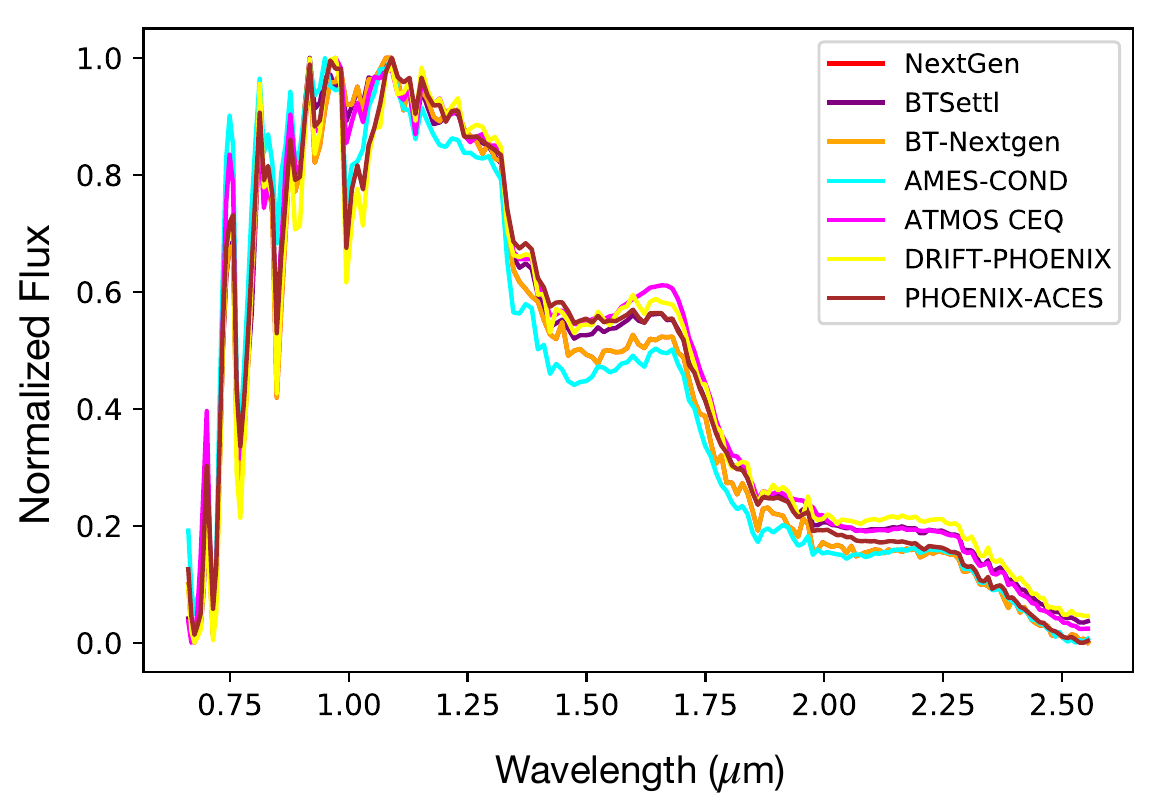}
    \caption{
        Showing disagreements between spectra for different stellar model grids available to the community for $T_\mathrm{eff} = 3000$\,K, $\log g=5.0$, and [Fe/H]$=0.0$. Here we see a wide range of differences in spectral shape that may be attributed to model assumptions or opacity database differences.
        \label{fig:modelcomp}
    }
\end{figure}

\begin{figure}[!tbp]
    \includegraphics[width=\columnwidth, height=\textheight, keepaspectratio]{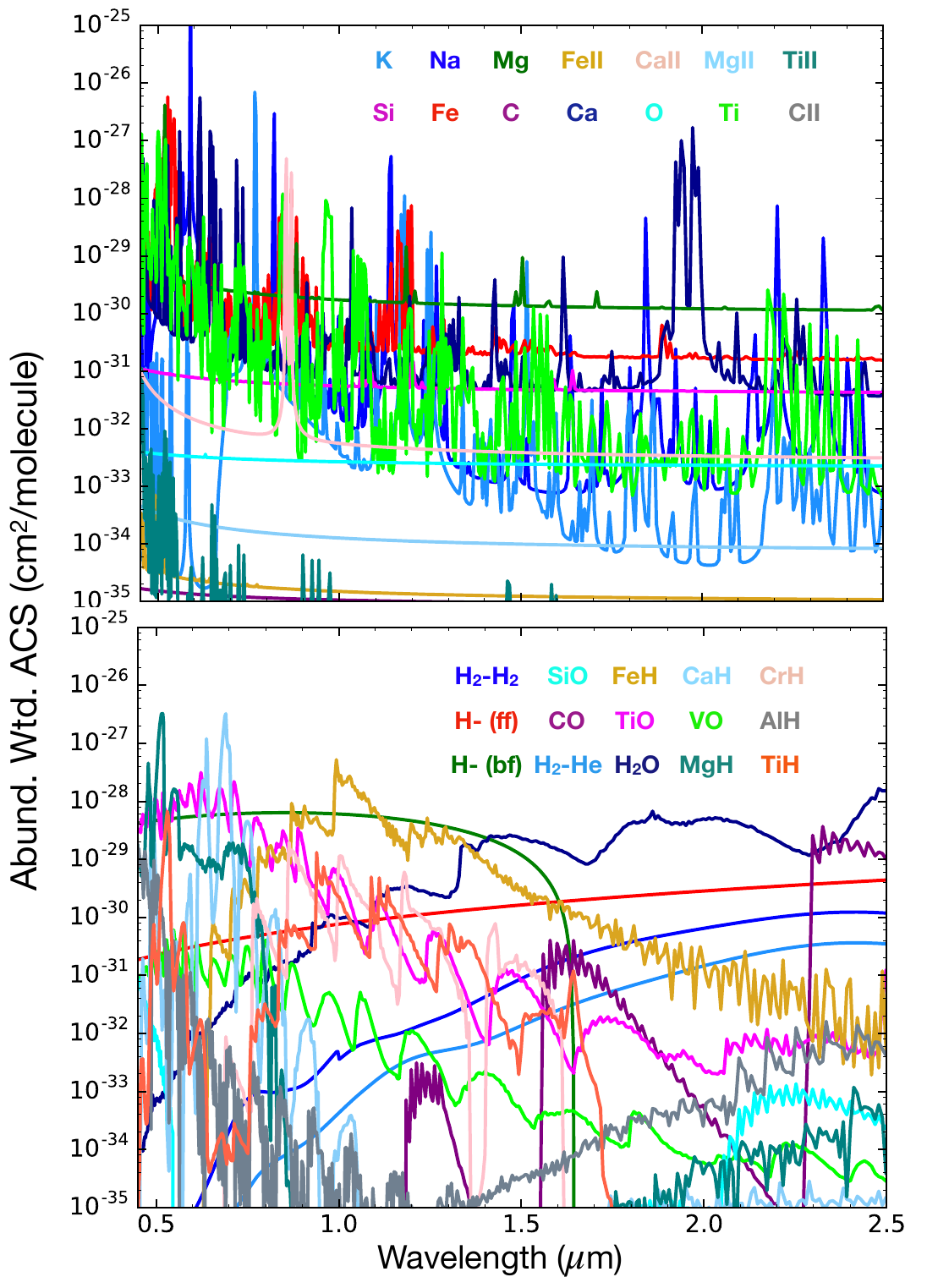}
    \caption{Abundance weighted absorption cross-sections (ACS) of atomic (top) and molecular (bottom) species included in our model grid from \textit{EXOPLINES} \citep{gharib2021exoplines} at a temperature of 3200 K and a pressure of 1bar covering a fiducial wavelength grid spanning 0.2--2.5$\mu$m at a resolution $\lambda$/$\Delta\lambda$ = 250.
    \label{opacplot}}
        
\end{figure}

Fundamental properties of cool stars include effective temperature, gravity, radius, and intrinsic elemental abundances. Processes of stellar formation, dynamics, the formation of molecules, atmospheric opacities, and energy balance are all dependent on accurate knowledge of these fundamental parameters. Previous works have focused on deriving relative elemental abundances from looking at specific lines (e.g. C, O, Fe, etc) (\citealt{chambers2010stellar,teske2014c,teske2016curious,ramirez2011elemental,ramirez2014chemical,melendez2009peculiar,fischer2005planet,brewer2015accurate}), which has worked adequately for FGK stars. Moreover, classic methods using synthetic spectral fitting techniques have generally focused on narrow spectral regions at higher resolutions (EW Method, \citealt{marfil2021carmenes}), and this has been particularly problematic for characterizing sub-solar mass stars. Due to their cooler effective temperatures and strategic placement between the main-sequence-FGK stars and brown dwarf, M dwarf atmospheres are plagued with multiple molecules and are known to have significant regions with turbulent convective interiors \citep{osterbrock1953internal}. These effects manifest in the form of broadband absorption features, thereby influencing the overall spectral energy distributions and producing a pseudo-continuum effect that varies with chemical abundances \citep{veyette2016physical}. Model grids assuming self-consistent one-dimensional radiative-convective equilibrium conditions available to the community today (\citealt{allard2000spherically,tsuji1996dust,tsuji2002water,kurucz2011including,husser2013,gustafsson2008grid}) show a wide variation in spectral shape for the same star (see Figure \ref{fig:modelcomp} as an example, for T$_{eff}$=3000 K, log$g$=5.0, solar metallicity star), possibly due to differences in atomic and molecular line lists used or the set of model assumptions incorporated.
Some of these differences could be further exacerbated by various stellar atmosphere effects that are not yet well understood; such as photospheric heterogeneity in active, fully/partially convective M dwarfs or dust physics \citep{apai2018understanding,patience2012spectroscopy}. \citealt{passegger2022metallicities} (and references therein) show broad inconsistencies when comparing fundamental parameters of M dwarfs derived from different methods such as pseudo-Equivalent Width measurements, spectral fitting techniques, empirical calibrations or machine learning methods relative to observations, despite employing standardization techniques such as consistent synthetic spectral models, wavelength ranges from consistent sets of observations, flux normalizations, dimensionality reduction in spectral fitting by fixing specific parameters etc. In fact, \citealt{passegger2022metallicities} found that different methods employed to characterize M dwarf abundances show consistency with the overall literature means of benchmark M dwarfs when used without these standardization procedures, within their native frameworks.

Naturally, huge implications could potentially arise due to these unresolved differences when it comes to modeling M dwarf atmospheres; such as (1) improper characterization of terrestrial exoplanets around M dwarfs with unocculted spots and sporadic photospheric activity would continue to persist despite correcting for these stellar effects even within a retrieval framework on high-fidelity JWST transmission spectra if the host star is not well understood \citep{iyer2020influence}, or (2) systematic uncertainties in analyzing population trends of spectral fitting derived effective temperature, metallicity and alpha enhancement parameters could persist and propagate into galactic chemodynamical evolution studies as a result of model deficiencies, especially for early M-s (T$_{eff}$ $>$ 3500 K) \citep{hejazi2022chemical}, or (3) inaccurate estimates of M dwarf atmospheric C/O ratios---particularly wide-separated primary companions of cooler brown dwarfs (such as Ross 19 AB, \citealt{schneider2021ross}) or exoplanets companions could result in incorrect conclusions about formation mechanisms of these sub-stellar mass objects \citep{moses2012chemical,madhusudhan2012c,mordasini2016imprint}. These are only a few consequences that plague our understanding of the complete picture concerning such stellar systems. Improper understanding of M dwarfs would therefore pose a significant setback in answering some of the most fundamental questions about our Universe, thereby emphasizing the crucial nature of efforts aimed at characterizing these stars. 

In light of these, we attempt to remedy a few of the challenges by: (1) Developing a new stellar synthesis model and (2) validating our model derived stellar fundamental parameters with benchmark M dwarf targets.
In section \ref{sec:newmod}, we introduce our new stellar atmosphere model grid, followed by a description of the low-resolution data used for this analysis. Then; we present our error mitigation approach using $Starfish$ \citep{czekala_github,ian_czekala_2018_2221006}, a Bayesian framework with Gaussian process kernel accounting for grid-interpolation uncertainties, data systematics, or model uncertainties and our robust estimates to fundamental properties of M dwarfs including T$_{eff}$, log$g$, Metallicity, C/O and Radius. In section \ref{results} we validate our grid model retrieved metallicity values with empirically measured [M/H] for widely separated M+G Binary dwarfs (hereafter, WBS) \citep{mann2013metal} as well as our retrieved radii with those measured with interferometric angular diameters (hereafter, IS) \citep{boyajian2012stellar}, using the SpeX, SNIFS, HST STIS observations from \citealt{mann2013metal,mann2015constrain} and references therein, all convolved to low spectral resolution regime (R$\sim$ 120). In section \ref{sec:discussions}, we show the influence of including additional parameterizations to retrieve for stellar surface photospheric heterogeneity as well as model assumed mixing-length convection parameter and their influence on retrieved metallicities, effective temperatures and radii. Finally, we end with a brief summary and implications of our work in section \ref{sec:summary}.

\begin{figure*}[!tbp]
    \includegraphics[width=\textwidth, height=\textheight, keepaspectratio]{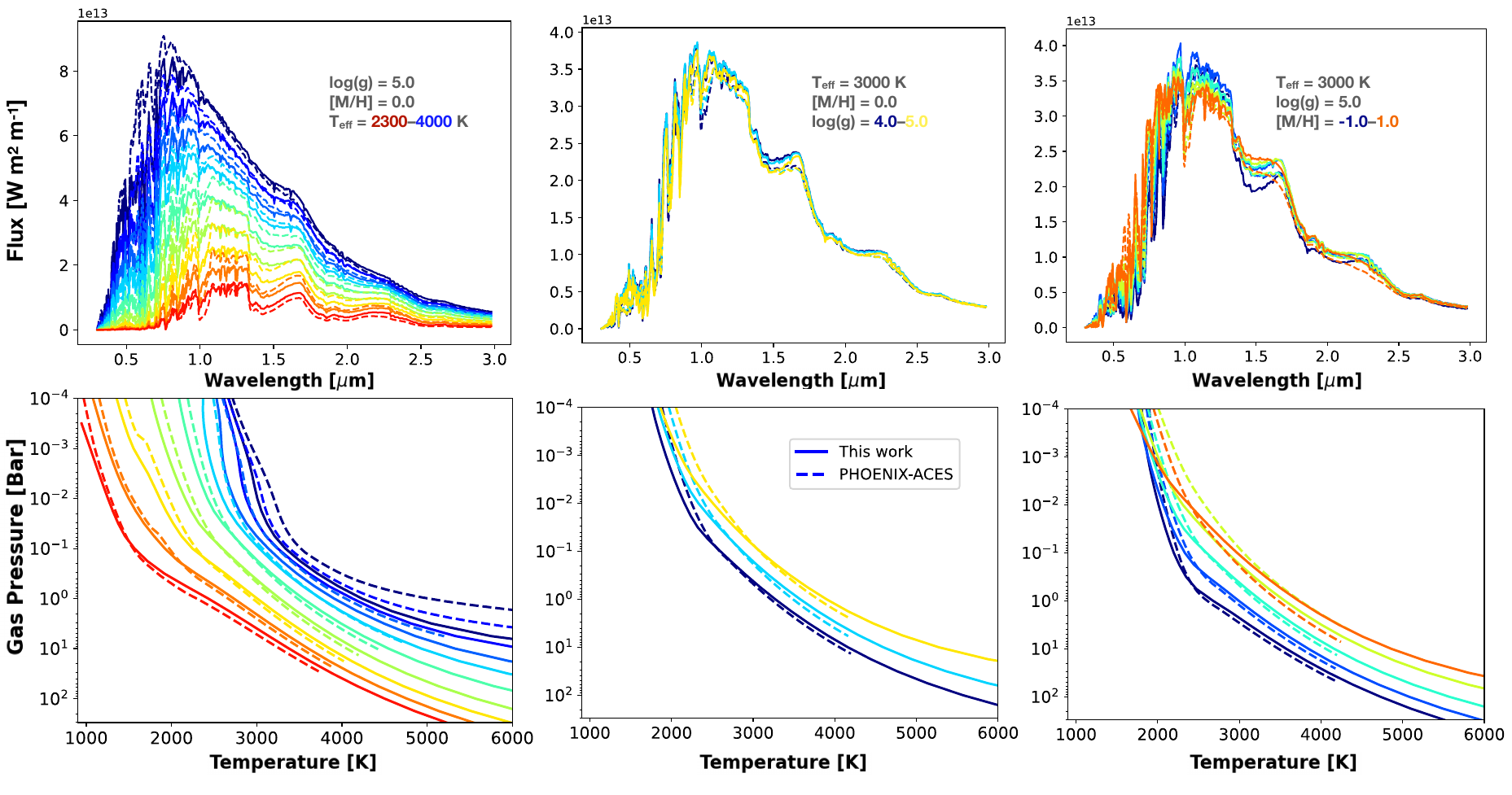}
    \caption{
        Comparison of our model grid spectra (solid lines) to PHOENIX-ACES high resolution model (dashed lines, smoothed to R$~$250) \citep{husser2013} in the top panel varying (left) T$_{eff}$ with fixed log$g$ = 5.0 and [Fe/H],[M/H] = 0.0, (center) varying log$g$ with fixed T$_{eff}$ = 3000 K, [Fe/H],[M/H] = 0.0, and (right) varying [Fe/H],[M/H] with fixed T$_{eff}$ = 3000 K and log$g$ = 5.0. Each bottom panel corresponds to their respective atmosphere structure profile. Here we see that our model grid spectra are in overall agreement with the latest PHOENIX-ACES grid, with most differences arising in varying the stellar metallicity and effective temperature. We include the most up-to-date molecular line-list data \citep{gharib2021exoplines} to compute opacities for M dwarf atmospheres, possibly explaining the main differences.
        \label{oursvsphx}
    }
\end{figure*}

\begin{table*}
\renewcommand{\thetable}{\arabic{table}}
\centering
\caption{Targets used for this analysis: 10 widely separated M+G Binaries (WBS), and 10 targets with interferometrically measured angular diameters (IS).Listed here are also their respective H-band magnitudes, distances and spectral type. The WBS spectra were taken from SNIFS \citep{lantz2004snifs,aldering2002overview} in optical and SpeX in IR \citep{rayner2003} by \citealt{mann2013metal} and the IS spectra were taken from SNIFS \citep{lantz2004snifs,aldering2002overview} + STIS \citep{bohlin2001,bohlin2004} in optical and SpeX in IR \citep{rayner2003} by \citealt{mann2015constrain}.}\label{data}
\begin{tabular}{|l | l | l | l |}
\hline
Target  & H-band Magnitude$^{a,b}$ & Distance$^c$ (pc) & spT$^{d,e}$\\
\hline
Wide Binary Sample (WBS) & & & \\
\hline
\decimals
\hline
NLTT 11270 & 10.257 $\pm$ 0.021 & 69.1884 $\pm$ 0.1565 & M0.4\\
NLTT 11500 & 8.749 $\pm$ 0.026 & 36.2962 $\pm$ 0.0829 & M1.8\\
NLTT 3725 & 8.927 $\pm$ 0.023 & 20.4716 $\pm$ 0.0479 & M3.9\\
NLTT 738  & 10.231 $\pm$ 0.022  & 76.7054 $\pm$ 0.2795 & M2.2\\
NLTT 8787 & 6.080 $\pm$ 0.038 & 11.1998 $\pm$ 0.0192 & M1.6\\
Gl 100 &  8.571  $\pm$ 0.029 & 19.6610 $\pm$ 0.0210 & M2.5\\
Gl 105 & 6.793 $\pm$ 0.038 & 7.2221 $\pm$ 0.0046 & M3.5 \\
Gl 118.2 & 8.935 $\pm$ 0.021 & 22.5773 $\pm$ 0.0325 & M3.8 \\
Gl 81.1 & 7.763 $\pm$ 0.021 & 33.5565 $\pm$ 0.0458 & M0.1\\
NLTT 10349 & 11.043 $\pm$ 0.023 & 77.6536 $\pm$ 0.1960 & M1.0\\
\hline
Interferometric Targets (IS)  &  &  &\\
\hline
Gl 436 & 6.319 $\pm$ 0.023 & 9.7559 $\pm$ 0.0089 & M2.8\\
Gl 880 & 4.800 $\pm$ 0.036 & 6.8676 $\pm$ 0.0018 & M1.5\\
Gl 699 & 4.83 $\pm$ 0.03 & 1.8266 $\pm$ 0.0010 & M4.2\\
Gl 526 & 4.775 $\pm$ 0.206 & 5.4353 $\pm$ 0.0015 & M1.4\\
Gl 687 & 4.77 $\pm$ 0.03 & 4.5500 $\pm$ 0.0007 & M3.2\\
Gl 887 & 3.61 $\pm$ 0.23 & 3.2871 $\pm$ 0.0005 & M1.1\\
Gl 411 & 3.640 $\pm$ 0.202 & 2.5461 $\pm$ 0.0002 & M1.9\\
Gl 581 & 6.095 $\pm$ 0.033 & 6.2992 $\pm$ 0.0020 & M3.2\\
Gl 725A & 4.741 $\pm$ 0.036 & 3.5218 $\pm$ 0.0008 & M3.0\\
Gl 725B & 5.197 $\pm$ 0.024 & 3.5228 $\pm$ 0.0013 & M3.5\\
\hline
\hline
\end{tabular}
\begin{center}
\begin{minipage}{15.5cm}
$^a$ H-band magnitude of 2MASS All Sky Catalog of Point Sourcces \citealt{catri2003}\\
$^b$ GAIA GR2 Collaboration \citealt{gaia2018}\\
$^c$ Vizier Online Data Catalog \citealt{ducati2002}\\
$^d$ \citealt{mann2013metal}
$^e$ \citealt{boyajian2012stellar}
\end{minipage}
\end{center}
\end{table*}

\section{NEW MODEL ATMOSPHERES} \label{sec:newmod}
\subsection{1-D Radiative Convective Equilibrium Model}
Our 1D radiative convective equilibrium model builds upon that from the ~ScCHIMERA~ tool primarily developed for extra-solar planet atmospheres described in  \citep{bonnefoy2018,piskorz2018, Arcangeli2018,Kreidberg2018,gharib2019influence,gharib2021exoplines,mansfield2021unique}. The model iteratively solves for  volume mixing ratio vertical profiles, cloud/condensate properties, thermal structure, and the top-of-atmosphere disk integrated stellar spectrum. We then generate a grid of model spectra and atmospheres\footnote{Models will be available after publication acceptance}, with components of our model pipeline as outlined below: 
\begin{itemize}
    \item \textbf{Radiative Transfer}: We follow the two-stream source function approximation as described in \citealt{toon1989rapid} and \citealt{petty2006first}. We compute radiative transport through the atmosphere constrained by hydrostatic equilibrium in a plane-parallel geometry, solving for the temperature-opacity combination resulting in net zero flux divergence across each atmosphere layer \citep{hubeny2017model,marley2014cool} using the Newton-Raphson iterative scheme described in \citealt{mckay1989thermal}.
    \item \textbf{Equation of State:} We use the NASA-CEA package \citep{Gordon1994} to compute equilibrium chemical abundances of molecules as well as neutral and singly ionized species. This package is able to compute equilibrium chemistry abundances for thousands of species, however we only use it to compute abundances for a subset of key molecules and elemental species that are characteristic of M dwarf atmospheres. 
    \item \textbf{Opacity Sources:} We include opacities relevant to cool stellar atmospheres including CH$_4$, CO, CO$_2$, NH$_3$, PH$_3$, HCN, C$_2$H$_2$, H$_2$S, H$_2$-H$_2$/He CIA \citep{Freedman2014,lupu_roxana_2022_6600976}, H$_2$O \citep{polyansky2018exomol}, metal hydrides TiH, FeH, CaH, CrH, AlH, SiH, MgH, and metal oxides TiO, VO, SiO (EXOPLINES \citealt{gharib2019influence,gharib2021exoplines,ehsan_gharib_nezhad_2021_4458189}). We also include neutral atomic lines Na, K (\citealt{allard2016k,allard2019new}, Fe I, Mg I, Ca I, CI, Si I, Ti I, O I, ionized atoms Fe II, Mg II, Ti II, C II, Ca II, (\citealt{kurucz1993atomic,kurucz2011including,piskunov1995vald}) and bound-bound and bound-free H$-$ continuum \citep{john1988continuous}. Our molecular opacities are continually updated based on the latest line-lists from ExoMol (\citealt{tennyson2016exomol,tennyson20202020}). Opacity functions can be dense due to numerous transitions as a function of temperature leading to millions of lines within any spectral window depending on the resultant resolution. Therefore; to surpass the computational burden caused by integrating over the opacity function in a tedious line-by-line manner, we use the correlated-k resort-rebin approach as described in \citealt{amundsen2017treatment}. Here, the pre-computed cross-sections are effectively down-sampled with line-by-line bins mapped onto a smooth cumulative distribution function of line strengths per bin, while keeping the ranking of line strengths consistent through the atmospheric column  (\citealt{grimm2015helios,marley1996atmospheric,burrows1997nongray,fortney2010transmission}). The line transmittance integral is then computed by performing a Gaussian quadrature sum using the cumulative distribution transformation, choosing 10s of quadrature points per spectral bin (we use about 20 points). This opacity transformation allows for faster computation of the transmission function for opacities without significant information loss (\citealt{marley1996atmospheric,burrows1997nongray,fortney2010transmission,showman2009atmospheric,line2013,line2014systematic,line2015uniform}). In Figure \ref{opacplot}, we show all molecular, atomic and continuum opacities included in our model suite at 3200 K temperature and 1 bar pressure. In fact, the choice of line list used for computing key M dwarf molecular opacities can show a difference in spectral shape over a factor of 2 and a difference in the thermal profile of up to 60 K \citep{gharib2021exoplines}. Therefore, it is absolutely crucial to use the most up-to-date line-lists with appropriate pressure broadening treatment (see \citealt{gharib2021exoplines}) particularly for key molecules that dominate M dwarf atmospheres such as metal oxides (TiO, VO, H$_2$O) and metal hydrides (MgH, CrH, FeH, CaH, SiH, TiH), which is the highlight of our synthesis model. 
    

  \item \textbf{Convection:} Our convection model follows mixing-length theory \citep{bohm1958wasserstoffkonvektionszone} to compute diffusion of heat through a turbulent medium. We solve for turbulent diffusivity and upward transport velocity using a non-zero convective heat flux in the adiabatic region of the atmosphere. For simplicity, we assume a mixing length parameter value of 1. We use the Schwarzchild criterion to assess whether conditions for convection are met and therefore solve for the convective flux through each strata and self-consistently update the thermal profile for regions with steep gradients. Mixing Length Theory also allows for a continuous transition from a fully radiative to fully convective solution providing better control over P-T profile convergence and permitting for detached convection zones \citep{hubeny2017model}.
  
  \item \textbf{Dust/Clouds:} Dust/cloud particles form when solid grains or liquid droplets condense as a result of the gas vapor pressure surpassing the local saturation vapor pressure \citep{ackerman2001precipitating,helling2008comparison}. To determine the opacity contributions from these condensates, the total absorption and scattering efficiencies are calculated given the particle size and indices of refraction of the condensate, in addition to assuming the distribution of particle sizes as a function of altitude in the form of a lognormal distribution following \citealt{ackerman2001precipitating,sharp2007atomic}. For the current grid model spectra, we assume the M dwarf atmospheres to be cloud-free, however this model capability as described is readily available to expand to late and/or low surface-gravity M dwarfs, including (but not limited to) Mg$_2$SiO$_4$, MgSiO$_3$, CaTiO$_3$, Na$_2$S, KCl, and Fe.
  
  \end{itemize}
  
In Figure \ref{oursvsphx}, we show slices of our model grid vs PHOENIX-ACES\footnote{\url{http://phoenix.astro.physik.uni-goettingen.de/}} \citep{husser2013} (smoothed to R$\sim$ 250) varying over T$_{eff}$, log$g$, and [M/H] which we define as sum of all elemental abundances relative to hydrogen, with the accepted solar value being 0.0 in log space. Overall, we demonstrate that our models are consistent when compared to the widely used stellar model grids that are available to the community today, with noticeable differences particularly for the greater than solar metallicity M dwarfs and early Ms $>$ 3500 K, a likely consequence of using the most recent line lists.


 \begin{figure*}[!tbp]
    \includegraphics[width=\textwidth]{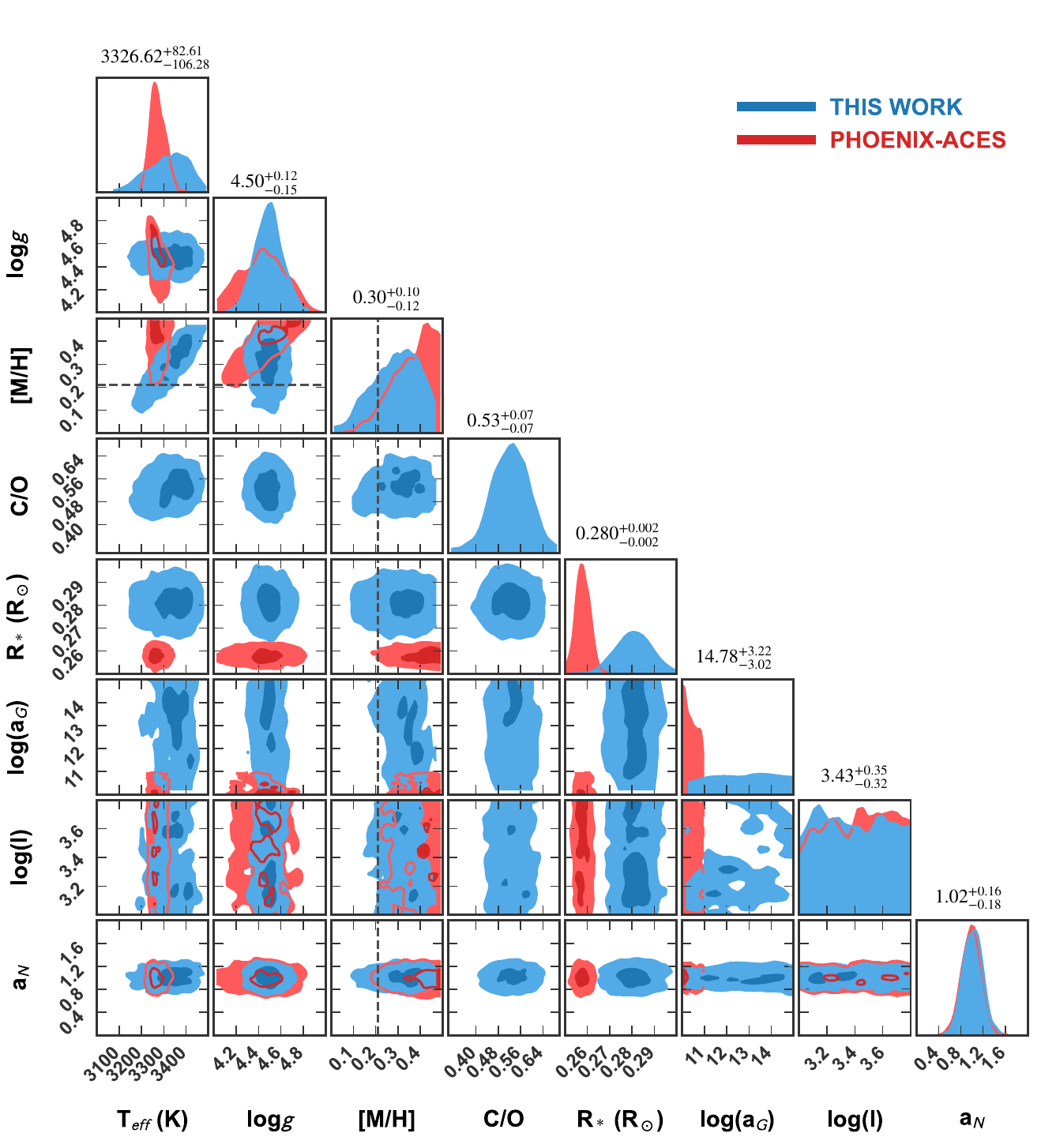}
    \caption{Grid Model fit retrieval results (blue) for NLTT 3725, a wide-separated binary G+M3.9 solving for fundamental stellar properties (T$_{eff}$, log$g$, [M/H], C/O and R$_*$) using our model grid compared to PHOENIX-ACES (red) \citep{husser2013}. Here we find that metallicity values retrieved using our new model grid provide a 1$\sigma$ improvement in the constraints when compared to empirically derived metallicity ([M/H] = 0.21) \citep{mann2013metal} shown in the dashed vertical line. In addition to physical stellar parameters we also show fits for $Starfish$ hyperparameters (log(aG), log(l), aN) which provide additional confidence to our values while incorporating proper sources of data and model uncertainties (see subsection \ref{subsec:starfish}). Our best-fit median and uncertainty values are provided in the inset text (see Table \ref{tab:bestfit}).} 
    \label{cornerplot}
  \end{figure*}

\subsection{Low-Resolution Data of Benchmark Systems}
We analyze M dwarf data in two groups: The first, wide-separated M+G binaries (WBS) with empirically calibrated metallicities ([M/H]) from the host stars \citep{mann2013metal}. The spectra obtained were observed with SpeX Spectrograph \citep{rayner2003} and SNIFS \citep{lantz2004snifs} covering a combined visible+NIR range spanning 0.32--2.4$\mu$m \citep{mann2013metal}. Both sets of observations were carried out by \citealt{mann2013metal} with the SpeX using NASA Infrared Telescope Facility (IRTF) on Mauna Kea with the SXD mode (0.3''x15'' slit at R$\sim$2000) and the SNIFS observations with University of Hawaii 2.2m telescope (with R$\sim$1300). \citealt{mann2013metal} then constructed the combined spectrum we use for our first group of M dwarf benchmark targets by stitching both data within the overlapping region of 0.81-0.96$\mu$m and performing appropriate RV corrections such that the offsets between both data are comparable to random errors in their RV measurements \citep{mann2013metal}.
For our second group, we focus on nearby-Ms from an  interferometry sample (IS) with measured angular diameters from \citealt{boyajian2012stellar}. For this group, we obtain data observed by \citealt{mann2015constrain} using SNIFS \citep{lantz2004snifs,aldering2002overview} and STIS \citep{bohlin2001,bohlin2004} on the \textit{Hubble Space Telescope} (HST) covering total optical bandpass of 0.32-0.97$\mu$m with R$\sim$1000 and NIR Spex \citep{rayner2003} covering 0.8-2.4$\mu$m at R$\sim$2000 \citep{mann2015constrain}. The SNIFS, SpeX and STIS data were again stitched, H-band flux calibrated and renormalized so as to minimize the relative offsets between the overlapping regions (see \citealt{mann2015constrain} for further details on the observations and data reduction procedures).

For both groups of M dwarfs, we choose to focus purely in the Optical/Near-Infrared wavelength range as it is ideal to infer molecular abundances and derive bulk chemical compositions \citep{line2015uniform} seamlessly. Furthermore, we also restrict ourselves only to low resolution regime as that provides a way to leverage broad-band molecular features over a wide wavelength range without the need to incorporate Non-LTE effects or Microturbulence that can otherwise persist when working with higher resolution spectra. Additionally, \citealt{jofre2014gaia} showed that varying instrument resolution resulted in metallicity differences of less than 0.05 dex for benchmark FGK stars they analyzed and less than 0.1 dex for stars below T$_{eff}$ of 5000 K. Such a paradigm shift to lower resolutions has also been demonstrated in galactic abundance synthesis models, and brown dwarfs (\citealt{ting2018,line2015uniform,line2017,zalesky2019,zhang2021,ZJ_paper2,zalesky2022uniform}). Our model grid spectra are computed at R$\sim$250 (see Figure \ref{oursvsphx}), however we smooth down both the data and model to R$\sim$ 120 for this analysis to match a generic SpeX Prism Library Spectra \citep{burgasser2014spex} as previously demonstrated for T dwarfs using $Starfish$ \citep{czekala2015starfish,czekala_github} by \citealt{zhang2021} and mainly to demonstrate the strength in our methodology in the low spectral resolution regime. 


\subsection{Grid-Model Interpolation routine}\label{subsec:starfish}
We use a Bayesian Inference tool $Starfish$\footnote{\url{https://github.com/Starfish-develop/Starfish}} \citep{czekala_github,ian_czekala_2018_2221006} as our primary grid model interpolation method as successfully used for M/G-type stars \citep{czekala_github,ian_czekala_2018_2221006,gully2017starfish}, benchmark brown dwarfs \citep{zhang2021,ZJ_paper2} and Protostars \citep{greene2018}. We first construct a spectral emulator matrix with $Starfish$, specifying the range of our grid of models: 2000 K $<=$ T$_{eff}$ $<=$ 4000 K, -1 $<=$ [M/H] $<=$ 1, 0.3 $<=$ C/O $<=$ 0.7, 4.0 $<=$ log$g$ $<=$ 5.5. Similar to \citealt{zhang2021}, we first reduce the spectral resolution of our model spectra for all grid points using a Gaussian kernel matching 0.''3 slit of SpeX Prism mode \citep{rayner2003} to be R$\sim$120. The convolved models are then used by the spectral emulator and deconstructed into eigenspectra using Principal Component Analysis. Within the emulator matrix, the entire grid is evaluated with a training set and a testing set and these are subsequently standardized by subtracting the mean spectrum over the entire grid as well as the standard deviation spectrum, along with associated weights in the form of PCA eigenspectra (\citealt{czekala_github,ian_czekala_2018_2221006,zhang2021}). Finally, for any given set of parameters spanning our grid, a reconstructed (or interpolated) spectrum is computed as a linear combination of the mean and standard deviation spectra and the associated eigenspectra (see equation 23 of \citealt{czekala_github,ian_czekala_2018_2221006} and equation 1 of \citealt{zhang2021}). Additionally, \citealt{zhang2021} demonstrated that reconstruction systematics (specifically for the Sonora-Bobcat models, \citealt{marley2021}) do not dominate the error budget of the forward modeling analysis. The Emulator Covariance Matrix employs a Gaussian Process method on the standardized weights for each eigenspectrum providing a probability distribution of the weights, a distribution of the interpolated spectra as well as the propagated interpolation uncertainties (see \citealt{czekala_github,ian_czekala_2018_2221006} and \citealt{zhang2021} for further details). 

The key feature of $Starfish$ is the construction of a composite covariance matrix, that accounts for the total error budget with parameter estimation as well as model evaluation. Components of the full covariance matrix are also inferred via retrievals (referred to as hyperparameters) to allow for realistic error analysis driven by the data. The first component of the covariance matrix; is a noise covariance matrix, which accounts for differences in the form of data-model uncertainties as well as observed flux uncertainties within and between wavelength pixels, inferred as parameter a$_N$. Second, a Gaussian process global covariance matrix, which uses a squared exponential kernel accounting for correlated residuals caused by oversampling the instrument line spread function (ILSF) or atmosphere model uncertainties as well as systematics (\citealt{czekala_github,ian_czekala_2018_2221006,zhang2021}), inferred as parameters log(a$_G$) and log($l$). This Gaussian Process kernel accounts for correlated residuals in the form of systematic differences arising from insufficient opacity data or other model deficiencies. With this method; the traditional gaussian likelihood function evaluation is modified to include a linear combination of the two components of the composite matrix following Equation 7 of \citealt{czekala_github,ian_czekala_2018_2221006} ultimately leaving us to solve for the following 8 parameters in our grid-model retrievals: T$_{eff}$, log$g$, [M/H], C/O, R$_*$, as physical parameters, and log(a$_G$), log($l$), a$_N$ as hyperparameters (see Figure \ref{cornerplot}). Note that we essentially fit for the (R$_*$/d)$^2$ scaling parameter as proxy for the radius R$_*$ constrained by distances measured from the \citealt{gaia2018}.

\subsection{Priors}
Our grid model fitting pipeline allows for choice of priors on both physical parameters as well as the hyperparameters. For the effective temperature, surface gravity, metallicity, carbon-to-oxygen ratio, and radius we use uniform priors spanning the range of the grid (2000 K $<=$ T$_{eff}$ $<=$ 4000 K, 4.0 $<=$ log$g$ $<=$ 5.0, -1 $<=$ [M/H] $<=$ 1, 0.3 $<=$ C/O $<=$ 0.7 and 0.01 R$\odot$ $<=$ R$_*$ $<=$ 0.8 R$\odot$). For the sake of efficiency, we truncate the grid ranges for each parameter axis informed by our linear interpolation routine (as done in \citealt{zalesky2019}) which precedes our $Starfish$ analysis, pertaining to each target in our sample. We found that this speeds up the $Starfish$ emulator matrix reconstruction for PCA training set over the grid points and its speed by $\sim$1 order of magnitude. Within the parameter ranges of our subgrid$-$specific to each target, we define physical priors to be uniform. For hyperparameters, we follow recommendations from \citealt{zhang2018} and \citealt{czekala_github,ian_czekala_2018_2221006} and use a Gaussian prior on $a_N$ with a mean of 1.0 and width of 0.25 over all positive values, which can sufficiently describe the ``Noise Covariance Matrix" accounting for covariance in data-model residuals within each pixel. We use uniform priors on both ``Global Covariance Matrix" parameters $l$ and log($aG$) to be between [200, 2000 x 5] km s$^{-1}$  and (-$\infty,\infty$) respectively for our stitched SNIFS+SpeX \citep{lantz2004snifs,rayner2003} spectra \citep{mann2013metal,mann2015constrain} which describes the behavior of the covariance values between two wavelengths, the SXD-mode line spread function, and the amplitude of the Gaussian Process Kernel to quantify correlated residuals possibly in the form of model systematics. While we did not rigorously compute the optimum auto-correlation wavelength to infer the appropriate range for the wavelength scale parameter ($l$) given our choice of data as done in \citealt{zhang2018}, we did however assume the average offset in the stitching of the two spectra (SNIFS+SpeX \citep{lantz2004snifs,rayner2003}) to be ranging between 30-50 
$\AA$ as indicated by \citealt{mann2013metal,mann2015constrain} (which corresponds to around 429--829 km s$^{-1}$ using equation A3 of \citealt{zhang2018}). We simply choose uniform priors on $l$ to comfortably accommodate this range as well as those prescribed by \citealt{zhang2018} when working with SpeX 0.''8 and 0.''5 slits. In a future analysis, fine tuning the inferences on the wavelength scale parameter ($l$) as well as the Gaussian Process Kernel amplitude log($aG$) will help in quantifying the impact of other possible sources of systematic noise that could influence the residuals \citep{czekala_github,ian_czekala_2018_2221006}.

\begin{figure*}[!tbp]
    \includegraphics[width=1.05\textwidth]{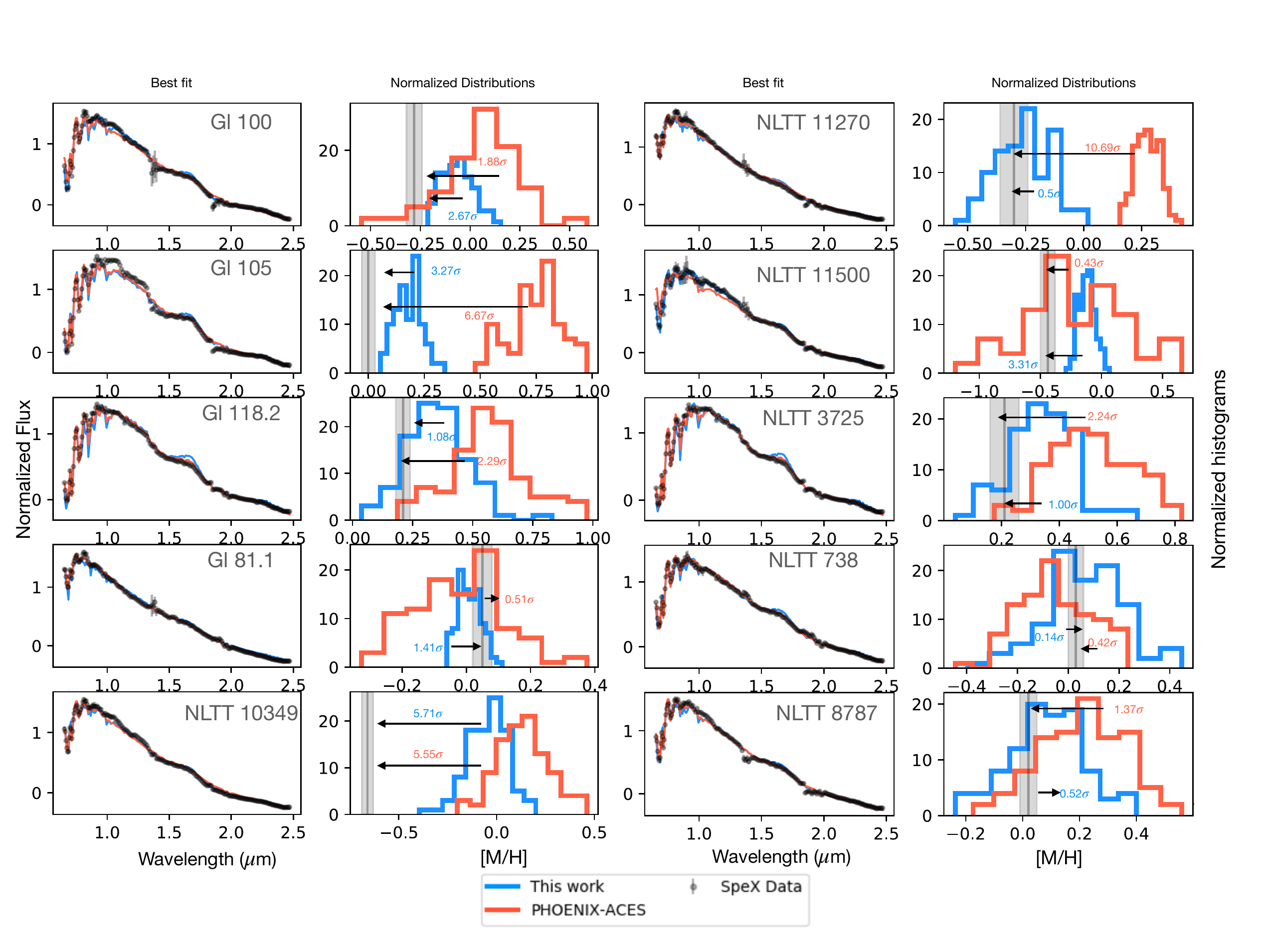}
    \centering
       \caption{Results for our WBS targets with empirically measured metallicity values \citep{mann2013metal}. Here in the first and third columns we show best fit spectra for the low-resolution data \citep{rayner2003,lantz2004snifs,mann2013metal} using our new M dwarf atmosphere model grid (blue) vs PHOENIX-ACES \citep{husser2013} (red) through our grid-model fitting pipeline. In columns two and four for each adjacent target we show posterior probability distributions of [M/H] from both fits, corresponding to the same color. In the inset of these histogram plots, we show in text the deviation (in units of $\sigma$ width of each histogram) of retrieved median [M/H] value compared to the empirical values and their uncertainty shown in gray vertical lines.} 
    \label{metallicityfits}
  \end{figure*}

\section{RESULTS}\label{results}
\begin{figure}[!tbp]
    \includegraphics[width=\columnwidth]{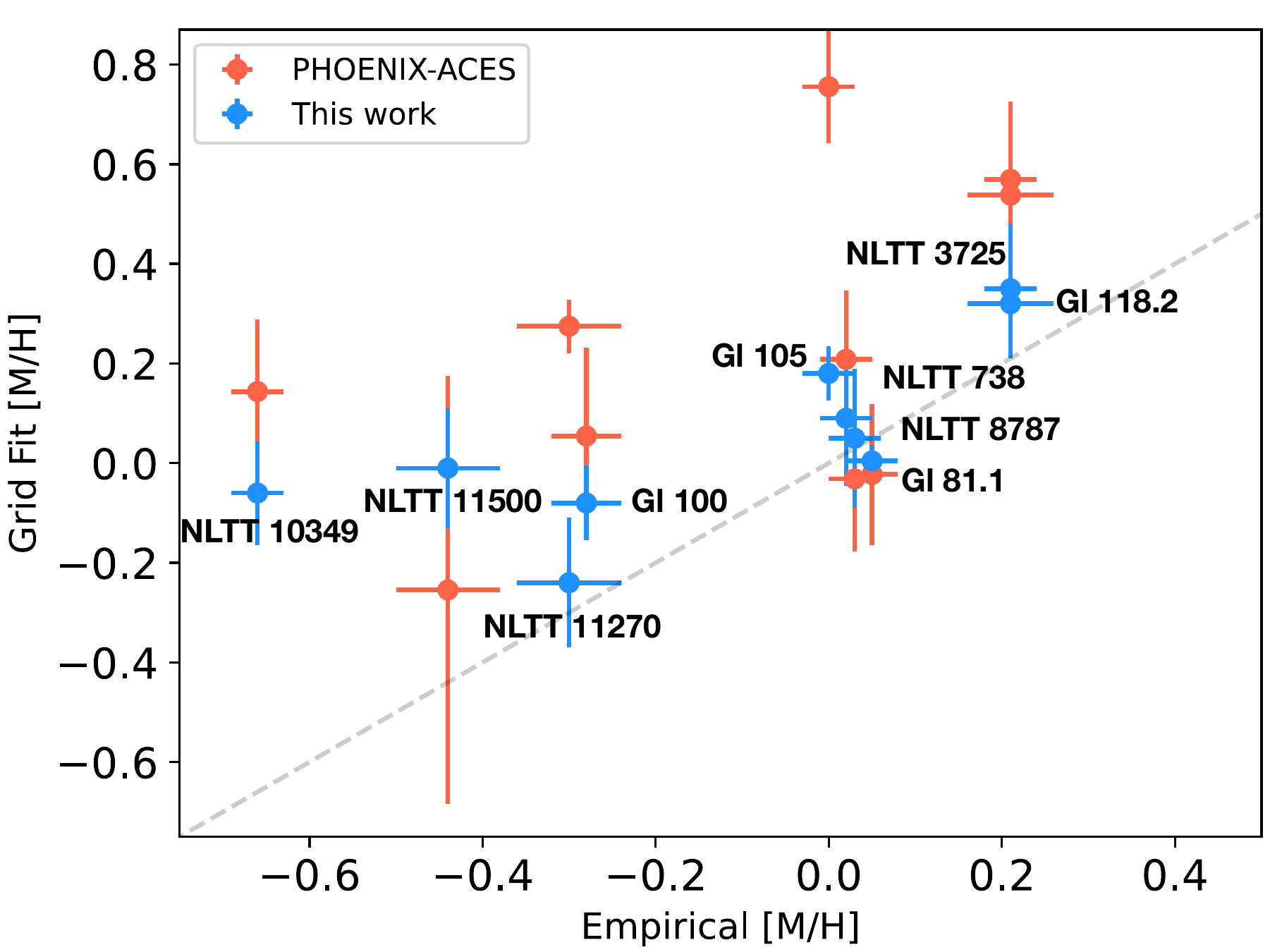}
    \caption{Comparing our metallicity values (blue) against empirically derived values \citep{mann2013metal}, also contrasting against values derived using the PHOENIX-ACES grid models (red) \citep{husser2013} within our grid model fitting routine. Our models fare well overall for all the M+G binary targets in our sample, improving constraints well within 1$\sigma$ for 60$\%$ of targets within our sample. The dashed grey line represents a slope of 1.} 
    \label{metphxcomp}
  \end{figure}

We first validate our grid models for both groups of M dwarf spectra: 10 Wide-separated M+G Binaries with empirically measured [M/H] (WBS) and 10 M dwarfs with interferometrically measured angular diameters (IS). All spectra are H-band flux calibrated and downgraded in resolution to R$\sim$120 for the entire SpeX SXD bandpass from 0.8 to 2.4 $\mu$m to focus solely in the low resolution regime. For all targets, we perform grid-model fits using the routine mentioned in section \ref{sec:newmod}, and retrieve fundamental stellar parameters (T$_{eff}$, log$g$, [M/H],C/O, and R$_*$), while also accounting for the $Starfish$ hyperparameters to accurately model errors. For our WBS targets we use data from SpeX + SNIFS that were stitched together in \citealt{mann2013metal}. We also perform fits using the PHOENIX-ACES \citep{husser2013} model grid through our fitting pipeline and we find our constraints to be overall improved when compared to the empirically measured values of [M/H] (see Figures \ref{cornerplot}, \ref{metallicityfits} and \ref{metphxcomp}). For 6 out of 10 targets our inferred metallicities are closer to the observational values in units of $\sigma$ of the retrieved histograms (see Figure \ref{metallicityfits}). For instance, the largest improvement is for M0.3 target NLTT 112700, where our [M/H] value is shifted 0.5$\sigma$ (the median lies within 1 $\sigma$ of the observations) versus PHOENIX-ACES \citep{husser2013} derived value through the $Starfish$ pipeline which is shifted over 10$\sigma$ from the empirical value \citep{mann2013metal}. Similarly, for M3.9 target Gl 105, we see about a 3$\sigma$ improvement using our grid-models. For Gl 118.2 (M3.8), NLTT 3725 (M3.9), and NLTT 8787 (M1.6), we collectively see about a 1 $\sigma$ improvement. For NLTT 738 (M2.2), we see no
significant difference in the retrieved metallicities; where both model values fall under 0.5$\sigma$ of the observations. However, for Gl 100 (M2.9), Gl 81.1 (M0.1), and NLTT 11500 (M1.8) we see that the PHOENIX-ACES models fare better. For two of the targets in this group (namely NLTT 10349 and NLTT 11500), we perform additional analyses in the next section to understand effects of photospheric activity that could perhaps improve model derived [M/H] values. 

While the overall fits using both our model grid as well as PHOENIX-ACES \citep{husser2013} perform well where each target best-fit residual is roughly $\sim$20$\%$, our inferred [M/H] provide for tighter constraints, owing to the inclusion of up-to-date opacities especially with key molecules that dominate M dwarf atmospheres such as metal oxides (H$_2$O, TiO, VO) and metal hydrides (MgH, CaH, CrH, SiH, TiH, AlH, and FeH) from the EXOPLINES database \citep{gharib2021exoplines}. 


  \begin{figure*}[!tbp]
    \includegraphics[width=1.03\textwidth]{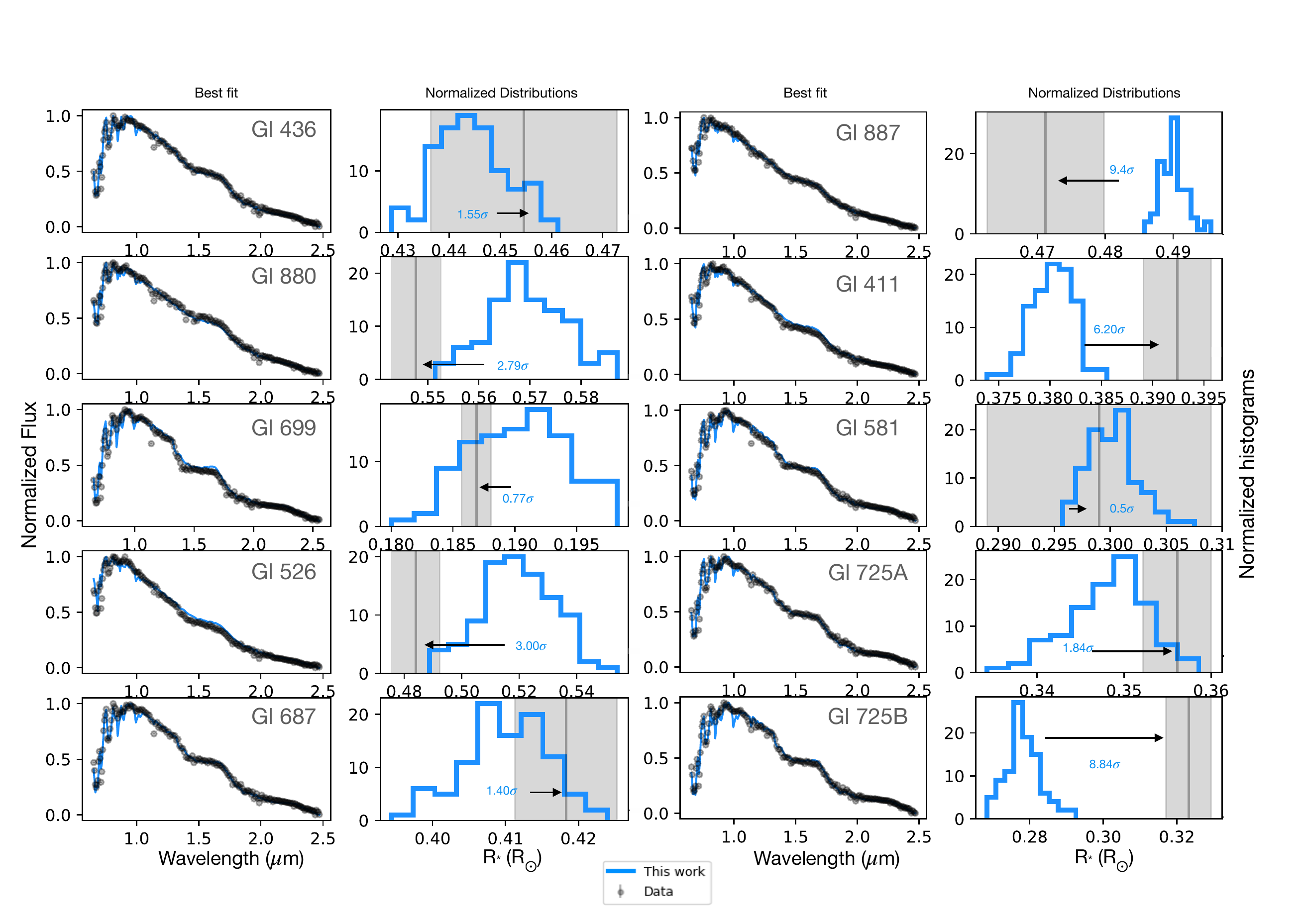}
    \centering
    \caption{Results for our IS targets with interferometrically measured angular diameters values \citep{boyajian2012stellar}. Here in the first and third columns we show best fit spectra for the low-resolution data \citep{rayner2003,lantz2004snifs,mann2015constrain} using our new M dwarf atmosphere model grid (blue) through our grid-model fitting pipeline. In columns two and four for each adjacent target we show posterior probability distributions of R$_*$ from both fits, corresponding to the same color. In the inset of these histogram plots, we show in text the deviation (in units of $\sigma$ width of each histogram) of retrieved median R$_*$ value compared to the the observed value and uncertainty shown in the gray vertical line \citep{boyajian2012stellar}.} 
    \label{radiusfits}
  \end{figure*}

\begin{figure*}[!tbp]
    \includegraphics[width=\textwidth,height=\textheight, keepaspectratio]{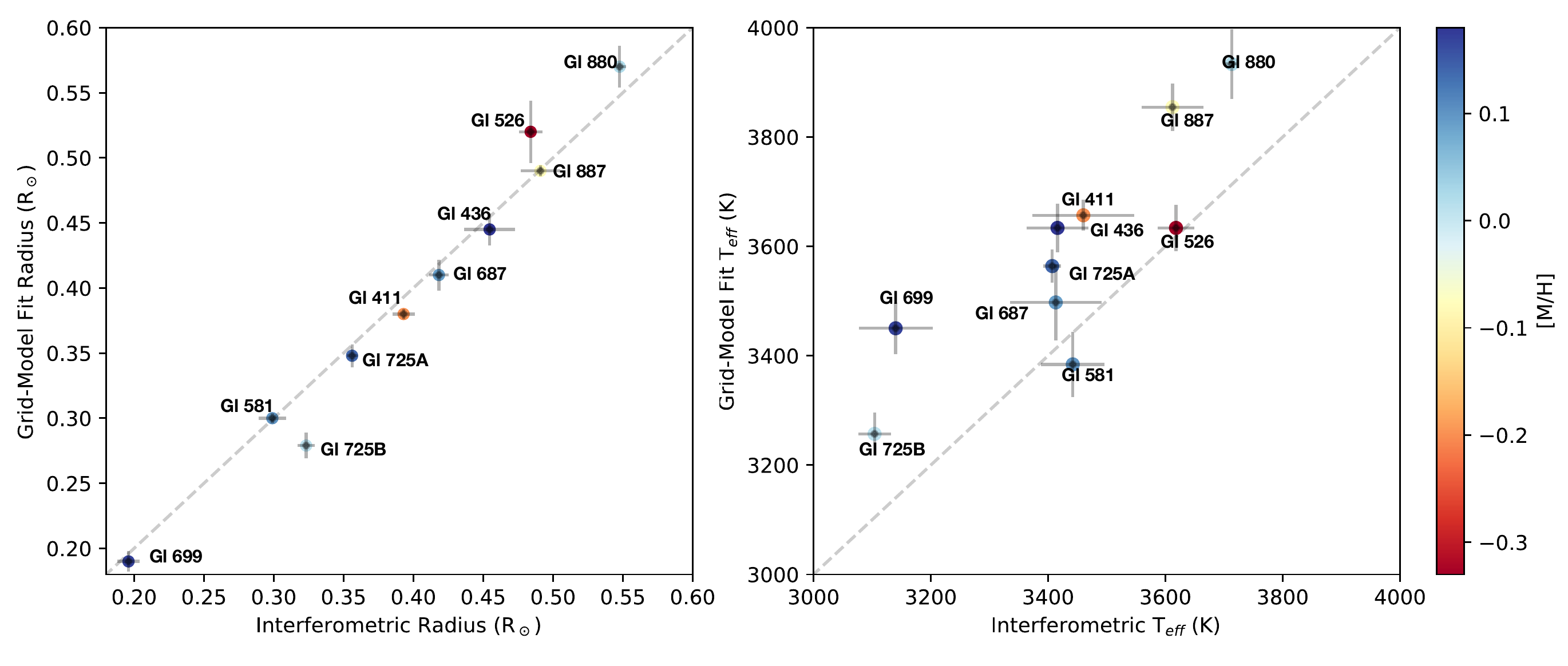}
    \caption{Comparing our retrieved radius (left) and effective temperatures (right) for the IS targets to those measured and derived interferometrically using the CHARA array \citep{boyajian2012stellar}. Here, we show that our radii values are quite consistent with the observations however our grid-model derived effective temperatures are overestimated. We also see here that there is no correlation with the derived metallicities of these stars (shown in the colorbar) to explain any of the deviations from observed values. We explore the effects of convective mixing length and photospheric heterogeneity to possibly explain some of these trends (see section \ref{sec:discussions}).}
    \label{radandTvals}
  \end{figure*}

For targets in our second sample group of M dwarf spectra: 10 Interferometry Sample (IS) with measured angular diameters \citep{boyajian2012stellar}, we focus to validate our model grid by inferring mainly the stellar radius, in addition to other fundamental stellar parameters such as T$_{eff}$, log$g$, [M/H] and C/O. We perform fits using the same grid-model fitting pipeline with our new stellar atmosphere models used for the WBS group (also described in section \ref{sec:newmod}). Here, we find that our grid-model retrieved radii are very close to interferometrically derived values \citep{boyajian2012stellar}. In Figure \ref{radiusfits} we see that 1 out of 10 targets; Gl 581 (M3.2) falls under 1$\sigma$, 3 out of 10 targets Gl 436 (M2.8), Gl 725A (M3.0), and Gl 687 (M3.2) fall within 2 $\sigma$ of the interferometrically retrieved R$_*$ and additional 2 targets Gl 526 (M1.4) and Gl 880 (M3.7) fall within 3$\sigma$. Note that the observational radii depicted in grey vertical lines in Figure \ref{radiusfits} are taken from \citealt{boyajian2012stellar} and references that lie within, however some targets have been provided with a weighted mean from multiple observations. For Gl 887 (M0.5) specifically, we find that our retrieved radius value is about 9$\sigma$ off from the mean interferometric radius of 0.4712 $\pm$ 0.0086 \citep{boyajian2012stellar} however we fare closer to 0.4910 $\pm$ 0.00140 acquired by \citealt{segransan2003first}, which is our adopted observational value in Figure \ref{radandTvals}.

\begin{figure}[!tbp]
    \includegraphics[width=\columnwidth]{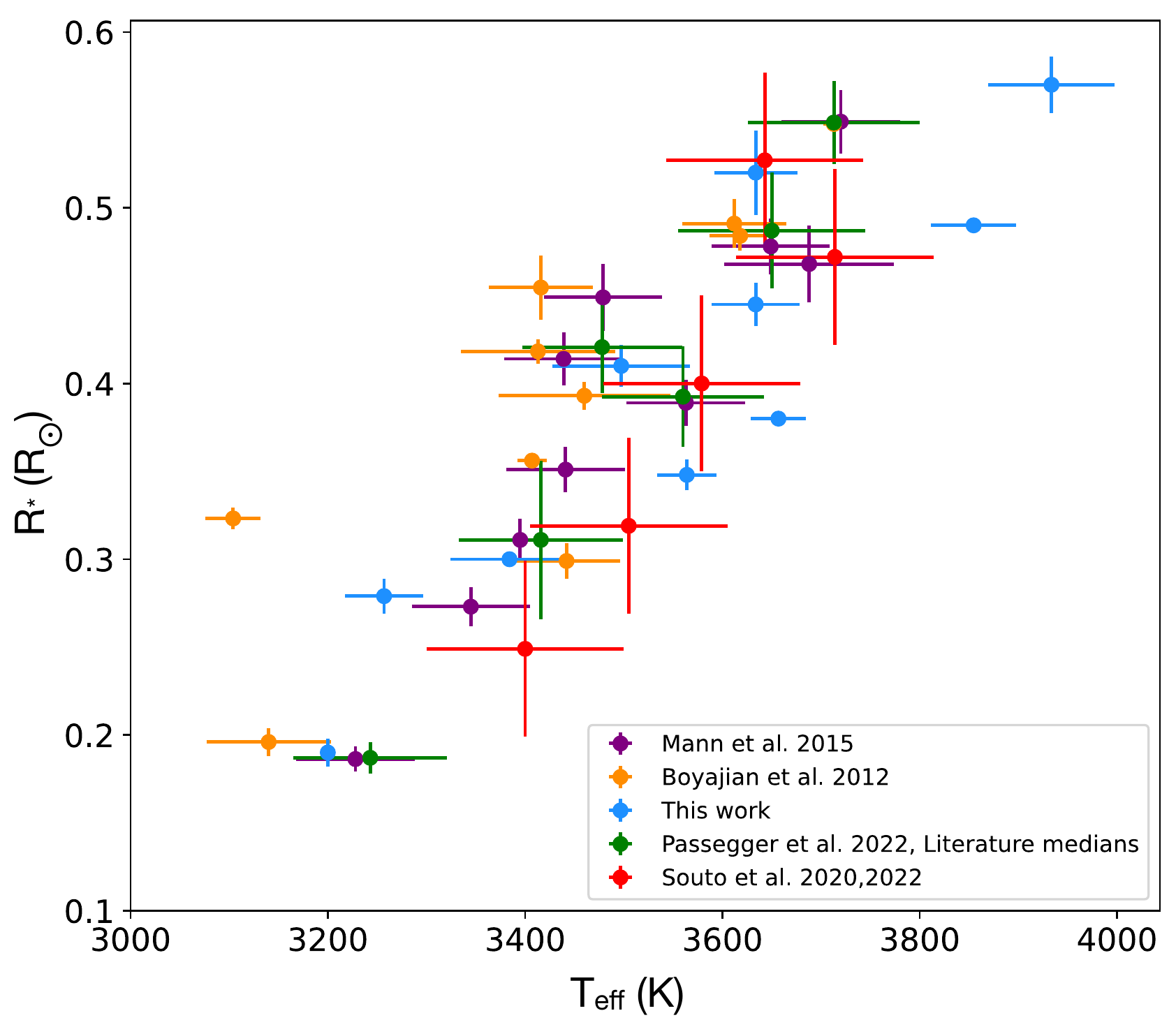}
    \caption{Comparing our model derived T$_{eff}$ and R$_*$ for the IS group targets. We overplot values derived from other works \citep{boyajian2012stellar,mann2015constrain,passegger2022metallicities,souto2020stellar,souto2022detailed} and show that our grid-model retrieved values are consistent within the scatter and a KS-2 sample test confirms they are drawn from the same parent distribution.}
    \label{ALLRADCOMP}
  \end{figure}

In addition to the retrieved radii, we also compare our grid-model retrieved effective temperatures of these targets to observations (\citealt{boyajian2012stellar} and references within). The overall trend is consistent without any significant correlations with [M/H] (shown in the colorbar) at least for our small sample of M dwarfs. We also find that our T$_{eff}$ values are systematically overestimated for a majority in our sample and R$_*$ are slightly underestimated for 5 out of 10 targets (similar to findings from \citealt{mann2015constrain}, and we discuss some possible explanations to these trends in the following section. We also show in Figure \ref{ALLRADCOMP} that our model retrieved T$_{eff}$ and R$_*$ are overall comparable to values acquired both from \citealt{mann2015constrain} using the BT-Settl version of PHOENIX \citep{allard2013bt} as well as interferometric observations done by \citealt{boyajian2012stellar} using the CHARA Array \citep{chara} with an average scatter of 4$\%$ and 5.2$\%$ respectively. Additionally, we also confirm that they are drawn from the same parent distribution with 95$\%$ and 99$\%$ confidence using a two-sample Kolmogorov-Smirnov test for CDFs of each dataset. We also overplot literature medians gathered from \citealt{passegger2022metallicities} for 6 targets that are in our IS sample and 5 targets from \citealt{souto2022detailed} for T$_{eff}$ and \citealt{souto2020stellar} for R$_*$ using spectral fitting of APOGEE High-resolution H-band spectra \citep{majewski2017apache} and confirm again with 99$\%$ and 95$\%$ confidence of being drawn from the same parent population as our derived radii and effective temperatures.

\begin{figure*}[!tbp]
    \includegraphics[width=0.75\textwidth]{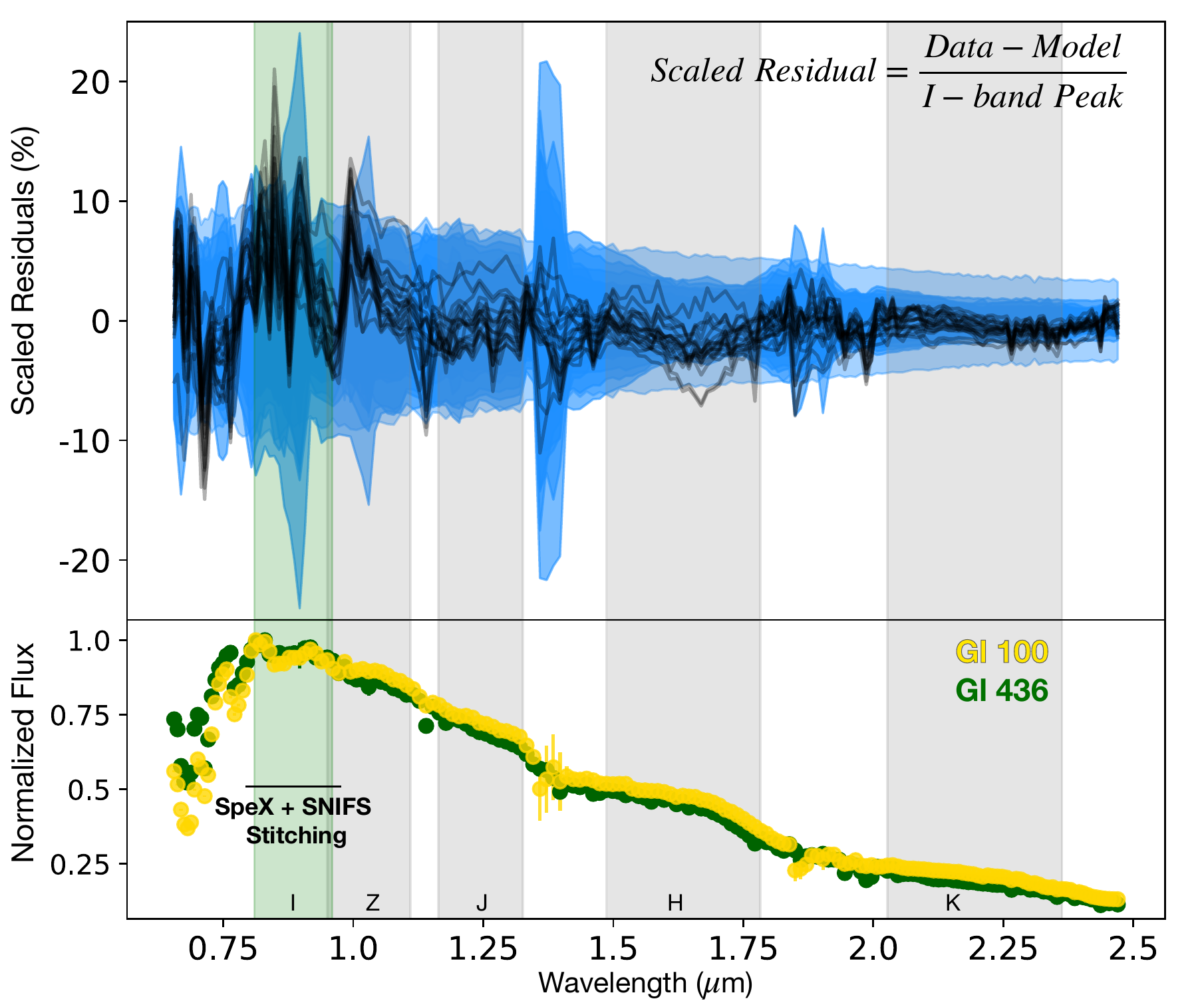}
    \centering
    \caption{(Top) Stacked residuals from our spectral grid model fits for all targets in our samples (black), including the widely-separated binary sample (WBS) and Interferometry sample (IS) with a total of 20 M dwarfs. The blue shaded regions are stacked 1$\sigma$ and 2$\sigma$ regions of 5000 random draws from the $Starfish$ \citep{czekala_github,ian_czekala_2018_2221006} covariance matrix for each target and (bottom) showing a spectrum of Gl 436 (dark green) and Gl 100 (yellow) for reference as representative of both groups within our sample, convolved to R~120. Both spectra and residuals are normalized by the observed peak I-band flux of each target. In the gray vertical shaded regions we show the SpeX Instrument \citep{rayner2003} defined filters of ZJHK bands and the green shaded region shows the SpeX+SNIFS overlap. We see that the largest deviations from our fits occur primarily at shorter wavelengths $<$ 1$\mu$m which could be due to a combination of data discrepancies as well as model deficiencies.}
    \label{starfisherror}
  \end{figure*}

For our entire combined sample of 20 M dwarfs, we also investigate spectral fitting residuals from $Starfish$ \citep{czekala_github,ian_czekala_2018_2221006} similar to \citealt{ZJ_paper2}. This allows us to explore model deficiencies or problematic observational contributions and how they affect spectral residuals as a function of wavelength. In Figure \ref{starfisherror}, we first stack all the data-model residuals (top panel, in black) from fits of each target normalized by their respective observed peak I-band flux. Overplotted in the same panel are stacked 1$\sigma$ and 2$\sigma$ dispersions over 5000 random draws from the $Starfish$ \citep{czekala_github,ian_czekala_2018_2221006} composite covariance matrix, also normalized by the I-band peak flux of each target. In the bottom panel of the same figure, we show one spectrum from each group of observations (WBS and IS) as representative of that group for reference. The residuals from all fits do not exceed 20$\%$ however, they show a significant wavelength-dependent trend with prominent features pointing to data stitching and telluric contamination at specific bands. We also see that the largest deviations appear at shorter wavelengths ($<$ 1$\mu$m), a region filled with bulky molecular opacities (see Figure \ref{opacplot}), showing possible sensitivity to model assumptions as well; such as considering a uniform non-heterogeneous photosphere or picking a solar calibrated mixing length value for convection for low mass stars, or lack of pressure broadening data of key M dwarf molecules and their influence on spectra (as discussed in \citealt{gharib2021exoplines}). While it appears difficult to decouple and quantify the individual contribution of model or data systematics, we are able to establish confidence in our retrieved fundamental stellar parameter values as we validate our models with benchmark M dwarfs while accounting for these combination error sources in a robust manner.

\section{DISCUSSION}\label{sec:discussions}
In this work, we introduce a 1-D self-consistent radiative-convective thermo-chemical equilibrium model atmosphere grid to characterize M dwarfs, focusing purely in low-resolution. We set out with the goal to infer fundamental stellar properties for these exoplanet hosting M dwarfs, and we show that not only are we able to acquire reasonable estimates for the stellar effective temperature, surface gravity, metallicity, and radius, but also extend to estimating their carbon-to-oxygen ratio with improved constraints. Knowing the C/O ratio of an exoplanet hosting star with increasing accuracy has vast implications in refining planet formation theories as well as estimating the C/O ratio of orbiting planets themselves, as they are necessary initial conditions for disk chemistry models \citep{fortney2012carbon}. 

Low mass stars such as M dwarfs in particular fall within the transition region from radiative to fully convective interiors around 0.28--0.33M$\odot$ \citep{chabrier1997structure}. It is around this region of the main-sequence stars where numerous challenges lie for both observational as well as theoretical efforts. We find for (WBS) a sample of 10 widely-separated binary companion M dwarfs, our retrieved bulk [M/H] values are consistent with empirically calibrated values (for 60$\%$ of our sample), however there are a few deviations that leave room for further exploration. We also find trends consistent with findings of \citealt{mann2015constrain,boyajian2012stellar} and \citealt{dieterich2021solar} for our (IS) targets with interferometric measurements where the T$_{eff}$ is overestimated for a majority in our sample and R$_*$ slightly underestimated for 5 out of 10 samples. We perform a few tests to further investigate these trends below. 

\subsection{Influence of Photospheric Heterogeneity}
First, we assess the influence of photospheric heterogeneity in the retrieved abundances. To answer this question, we include a parameterization \citep{apai2018understanding,rackham2017,rackham2018,pinhas2018,zhang2018,iyer2020influence} for the stellar spectrum to be a linear combination of two spectra--one representative of the calm photosphere and one representative of a cooler spot region on the photosphere that are both weighted by the fractional coverage in area of the spots: 

\begin{equation}
    F_* = F_{phot}(1-f_{spot}) + F_{spot}f_{spot}
\end{equation}

where F$_*$ is the final stellar model spectrum, F$_{phot}$ is the spectrum of the stellar photosphere region alone, F$_{spot}$ is the spectrum of the stellar spot region alone and finally f$_{spot}$ is the fractional coverage in area of the spots. Here, both F$_{phot}$ and F$_{spot}$ are model spectra as a function of their respective T$_{eff}$, log$g$, [M/H], C/O and scaled (R$_*$/d)$^2$ values evaluated (generated with the $Starfish$ interpolation emulator, see subsection \ref{subsec:starfish}) at each instantiation within our retrieval. All parameters are sampled as a single instance from their respective posteriors for each likelihood evaluation within our routine, except for the effective temperature T$_{eff}$ alone, which is drawn twice: one representing T$_{phot}$ for the temperature of the photosphere region, and T$_{spot}$ for the temperature of the spot region, to ultimately produce the final model spectrum F$_*$. Following this, we end up with three additional parameters for our grid-model retrievals namely, T$_{phot}$, T$_{spot}$ and f$_{spot}$. Both the temperature values are bound by uniform priors spanning the range of our model grid (except with the condition that the T$_{spot}$ is always cooler than the T$_{phot}$), and f$_{spot}$ is taken as a uniform prior between 0-1 as we do not have abundant \textit{a~priori} knowledge on physically reasonable spot covering fractions on M dwarf photospheres \citep{iyer2020influence}.

We perform retrievals including this ``mixed model'' parameterization for two targets with the largest $\sigma$ deviation from observations in Figure \ref{metallicityfits}, NLTT~10349 (M1.0) and NLTT~11500 (M1.8). In Figure \ref{metal_spot}, NLTT~10349 shows an improvement from 3$\sigma$ to 2$\sigma$ in our ability to constrain [M/H] relative to empirically derived value. At the same time, we find NLTT~11500 does not show any further improvements in [M/H] upon the inclusion of stellar spots. However, the retrieval does appear to be artificially pushing the spot covering fraction to be significantly high. We also see that the log$g$ posterior is driven up to higher values, necessitating further investigations while potentially exploiting the mass/metallicity correlation for a mid-M dwarf (see Figure \ref{metal_spot}, cont.).

We also test the influence of including stellar spot parameterization on the retrieved radii. For the case of Gl 725B (M3.5)--a target in our sample with the largest deviation from the interferometric measurements (IS), we find that inclusion of stellar heterogeneity ever so slightly improves accuracy in the retrieved radius. The retrieved effective temperature however is significantly more overestimated than before with an even larger deviation from interferometry, and this could perhaps be due to a simplified stellar spot parameterization causing degeneracies with other fundamental stellar properties, that don't always prove to be accurate for active M dwarfs \citep{iyer2020influence} (extra figures provided in our zenodo link). It is also well established by now that there is very little knowledge of physically reasonable coverage fractions of stellar photospheric heterogeneities as well as their distribution throughout the photosphere of an M dwarf, posing alongside numerous problems that exist in characterizing the activity environment of such stars \citep{rackham2022final}. Spots occur in stars primarily due to pockets of magnetic flux tubules stretching outwards beyond the photosphere suppressing the flux in that region as a result of convection processes that produce an overall cooler effect in this region. Therefore, efforts to model this more accurately including complementary efforts from the observational side could provide a great starting point for fine tuning priors on such correlated parameters, as well as improving parameterizations within a retrieval framework woud be of tremendous benefit.

\begin{figure*}[!h]
    \includegraphics[width=\textwidth]{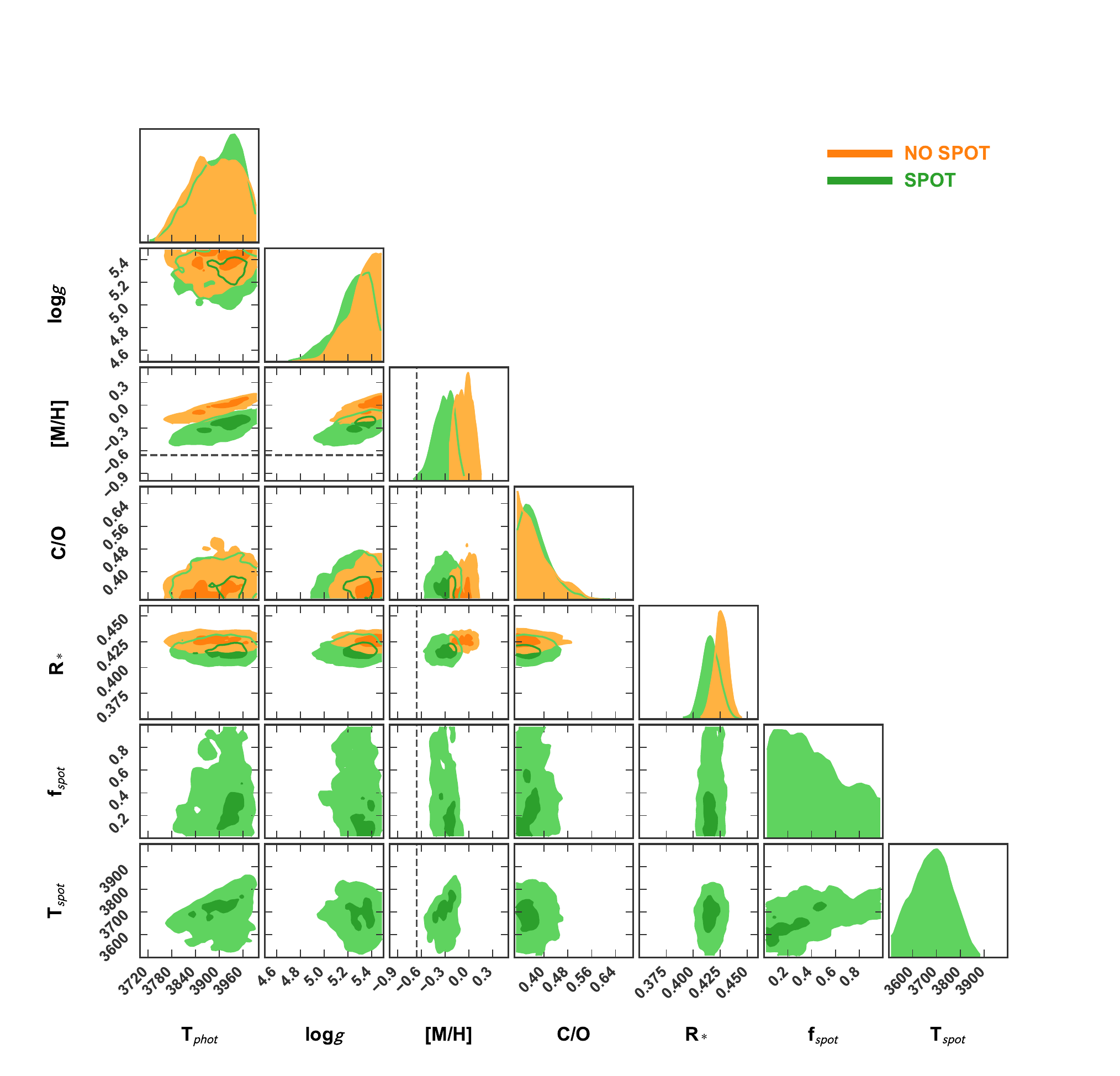}
    \centering
    \caption{Posterior probability distributions for two WBS (metallacity calibration) targets with largest deviations from empirical measurements (see Figure \ref{metallicityfits})--NLTT 10349 (Top) and NLTT 11500 (Bottom). Here we compare results of retrieved stellar properties with (green) and without (orange) inclusion of stellar surface heterogeneity parameterization. We find for NLTT 10349 we are able to further improve the [M/H] from 3--2$\sigma$ closer to the empirically derived [M/H] shown in inset as the black vertical dotted line. For NLTT 11500 however; there is no such change, except for the retrieved log$g$. The empirical [M/H] for this target is -0.44 \citep{mann2013metal}, which is beyond the range of the plot and therefore not shown.}
    \label{metal_spot}
  \end{figure*}
    \begin{figure*}[!tbp]
\renewcommand{\thefigure}{\arabic{figure}}
\addtocounter{figure}{-1}
    \includegraphics[width=\textwidth]{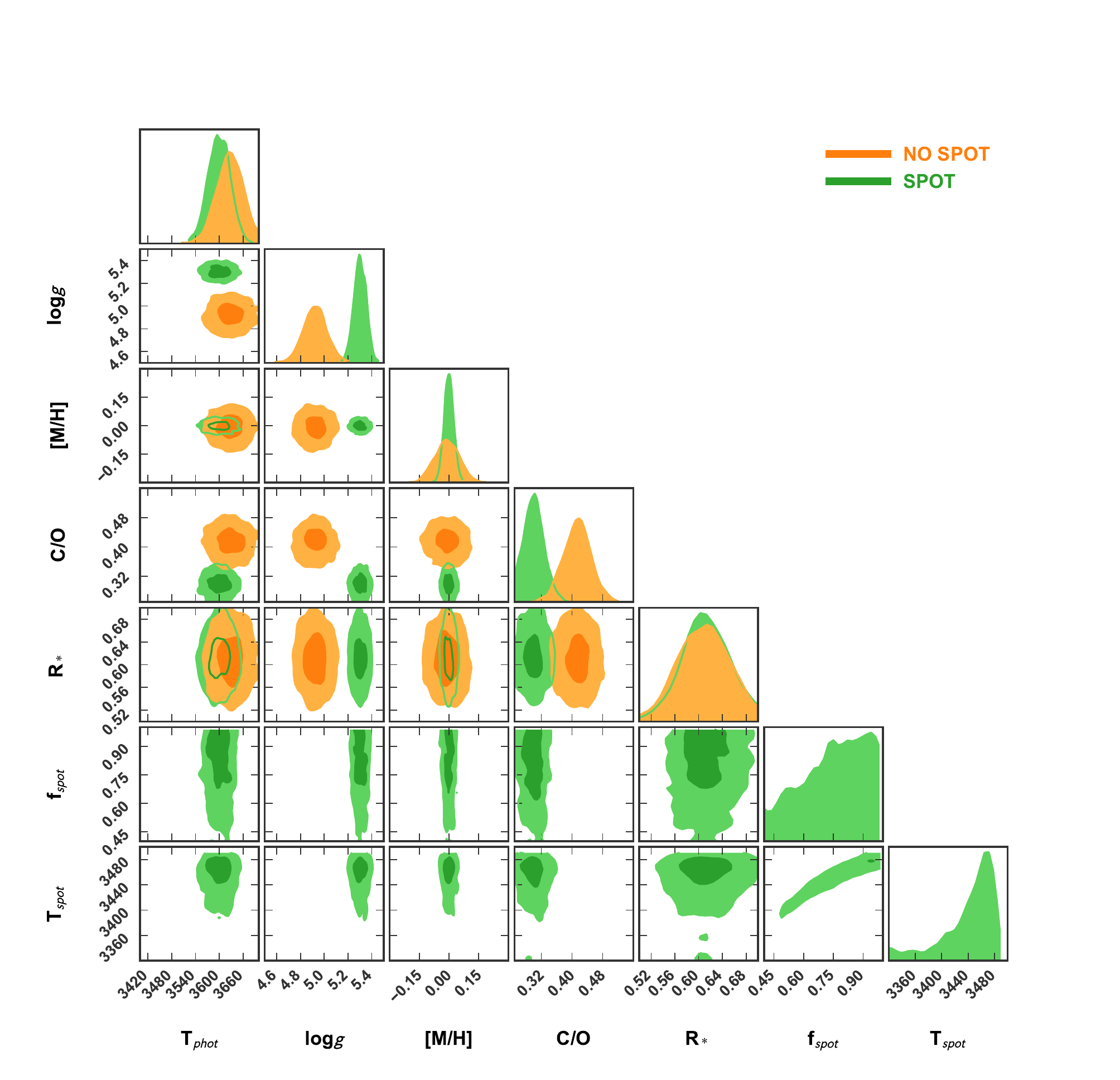}
    \centering
    \caption{(Continued)}
 \end{figure*}

\subsection{Influence of the Convective Mixing Length Parameter}
Here we explore the influence of the convective mixing length parameter \citep{bohm1958wasserstoffkonvektionszone} on our stellar atmosphere model retrieved values. We define the mixing length parameter as:

\begin{equation}
    \alpha = \frac{l_{mix}}{H_p}
\end{equation}

where $l_{mix}$ is the convective mixing length and H$_p$ is the pressure scale height, and the ratio $\alpha$ yielding a free parameter which is often set based on solar-calibrated values \citep{chabrier1997structure}. All our models in the grid as described in section \ref{sec:newmod} are calculated assuming $\alpha$~=~1 for simplicity. In Figure \ref{radandTvals}, we find that while most of our retrieved radii are consistent with the interferometry values they tend to be slightly underestimated (for at least 50$\%$ of our targets), with the most dramatic deviation seen for Gl 725B (M3.5). For a majority of these stars however, in the same figure (right panel) we find that their effective temperatures are overestimated relative to the interferometry values. To address this, we perform a test by changing the mixing length value to $\alpha$~=~0.5 and generate a mini-grid of models to assess whether it could improve our predictions and explain the systematic deviations seen in both stellar radii and effective temperature. We choose this value as recommended by \citealt{mann2015constrain} who found that reducing the mixing length parameter yielded a reduction in their T$_{eff}$ values of about 2$\%$ and increase in R$_*$ of 2.5$\%$ for low mass stars $<$
0.3M$\odot$. We also repeat this exercise with a value of $\alpha$~=~2 following \citealt{baraffe2015new} who primarily favor mixing length values above 1.6 for cooler and denser stars with larger convective envelopes. [M/H] could potentially be sensitive to choice of mixing length based on relevant model calibrations however, \citealt{song2020impact} find that the typical scatter in [Fe/H] values due to $\alpha$ are well within the uncertainty levels of metallicity itself. 

We also pick Gl 699 for this exercise (Barnard's star, which is known to have subsolar metallicity, [Fe/H] = -0.36 \citealt{rojas2012metallicity}), where our model appears to derive a super-solar [M/H] (0.17) and overestimated (3450 K) T$_{eff}$ (see Figure \ref{radandTvals}). In Figure \ref{gj725bspotcomp}, the model grid assuming $\alpha$=0.5 fares better relative to the $\alpha$=2 models in terms of the retrieved effective temperature, getting it closer to the interferometry value within 1$\sigma$ and 2$\sigma$ shift from the medians respectively for both Gl 725B and Gl 699. We do not see any changes in the retrieved radii for both cases however, the retrieved [M/H] value changes to better match the empirically derived value from \citealt{mann2015constrain} under 1$\sigma$ for Gl 725B and 2$\sigma$ for Gl 699 compared to observations. Particularly, the median retrieved [M/H] with the $\alpha$= 0.5 model results in a sub-solar metallicity for Barnard's star as expected. We also do not see a significant difference in the retrieved surface gravity parameter, log$g$, for both cases consistent with findings from \citealt{mann2015constrain} where the mixing length parameter did not appear to perturb derived masses significantly. Hence, for very low mass targets ($<$0.3M$_\odot$) such as Gl 725B and Gl 699, we find that effective temperature and metallicity are sensitive to the mixing length parameter assumed. Reducing $\alpha$ from the solar-calibrated value ($\alpha$=1.88, \citealt{mann2015constrain}) to a lower value ($\alpha$=0.5) is in essence equivalent to reducing the region of the convective envelope in the star, and increasing the radiative interior, at the same time cooling the surface temperature and influencing the molecular opacities in accordance. Consistently, we see a slight reduction in model derived effective temperatures and metallicities for both Gl 725B and Gl 699 that fall within the fully-convective regime ($<$0.3M$_\odot$, Figure \ref{gj725bspotcomp}). For a $>$0.3M$_\odot$ target Gl 436, reducing the mixing length parameter showed a dramatic reduction in effective temperature and metallicity, with no noteworthy change in the radius (see supplementary figure in zenodo link). In line with findings of \citealt{mann2015constrain} (see their Table 4), we also find that the relative difference in retrieved effective temperature due to reducing the convective parameter is likely larger for $>$0.3M$_\odot$ M dwarfs, however analyzing a larger sample and resulting population level trends will allow for a more thorough assessment.


We also investigated the effects of stellar spots for the case of Gl 725B alone, while keeping the mixing length parameter value constant at 1, and found that the model derived radius is slightly increased by an additional 1$\sigma$---thereby shifting the histogram closer to the observational value (supplementary figure provided in zenodo link). No significant changes occured in the other derived parameters. Therefore, we recommend that including mixed-models considering the effects of convection as well as stellar photospheric heterogeneity could be one way of understanding the nuances that go into influencing retrieved fundamental properties of M dwarfs, particularly when moving towards later spectral types.

  \begin{figure*}[!tbp]
    \includegraphics[width=\textwidth]{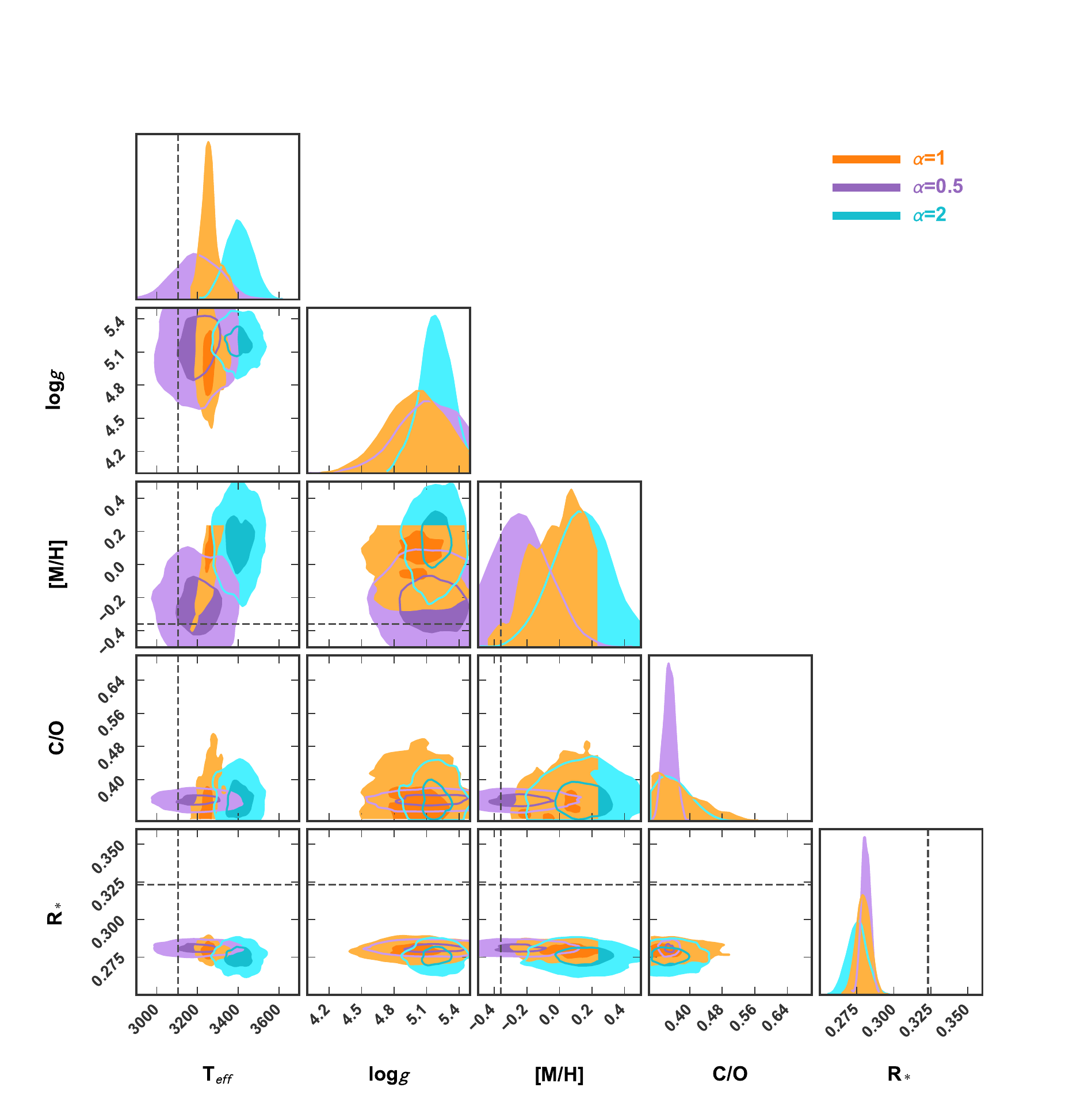}
    \centering
    \caption{Posterior probability distributions comparing fits for Gl 725B (Top) and Gl 699 (Bottom) using a mini-atmosphere grid generated for $\alpha$ = 0.5 (purple), $\alpha$ = 1 (orange) as well as $\alpha$ = 2 (cyan), both from our IS sample (radius calibration). Here we see that the retrieved radii are no different, however the retrieved T$_{eff}$ and [M/H] derived using the lower mixing length parameter ($\alpha$=0.5) model matches within 1$\sigma$ and 2$\sigma$ for Gl 725B and Gl 699 respectively, compared to observations \citep{boyajian2012stellar}. At least for these two cases, it appears that a smaller mixing length parameter $\alpha$ seems to be appropriate.}
    \label{gj725bspotcomp}
  \end{figure*}
    \begin{figure*}[!tbp]
\renewcommand{\thefigure}{\arabic{figure}}
\addtocounter{figure}{-1}
    \includegraphics[width=\textwidth]{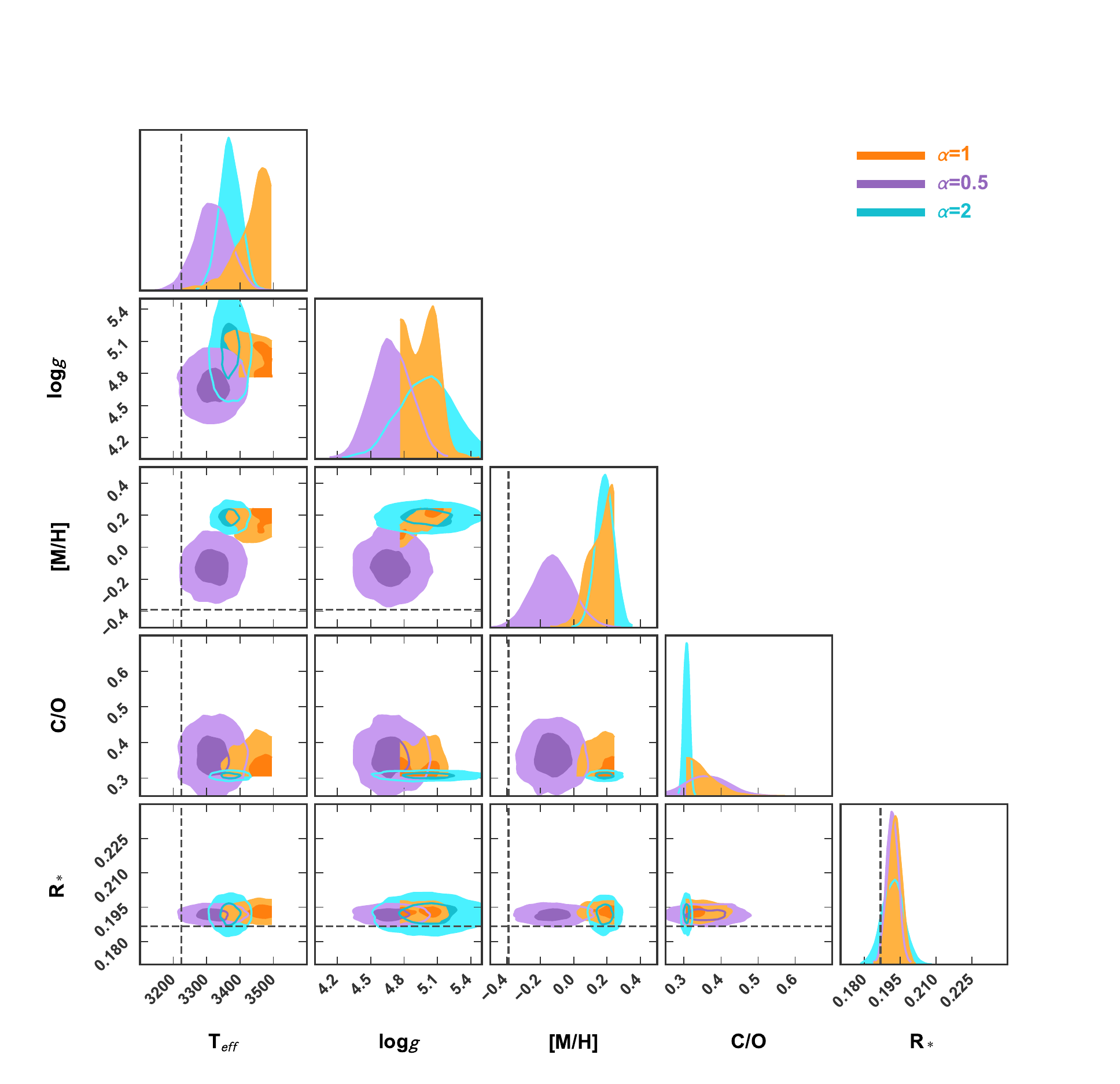}
    \centering
    \caption{(Continued)}
   \end{figure*}

\subsection{Carbon-to-Oxygen Ratio of M dwarfs}
\begin{figure}[!tbp]
    \includegraphics[width=\columnwidth]{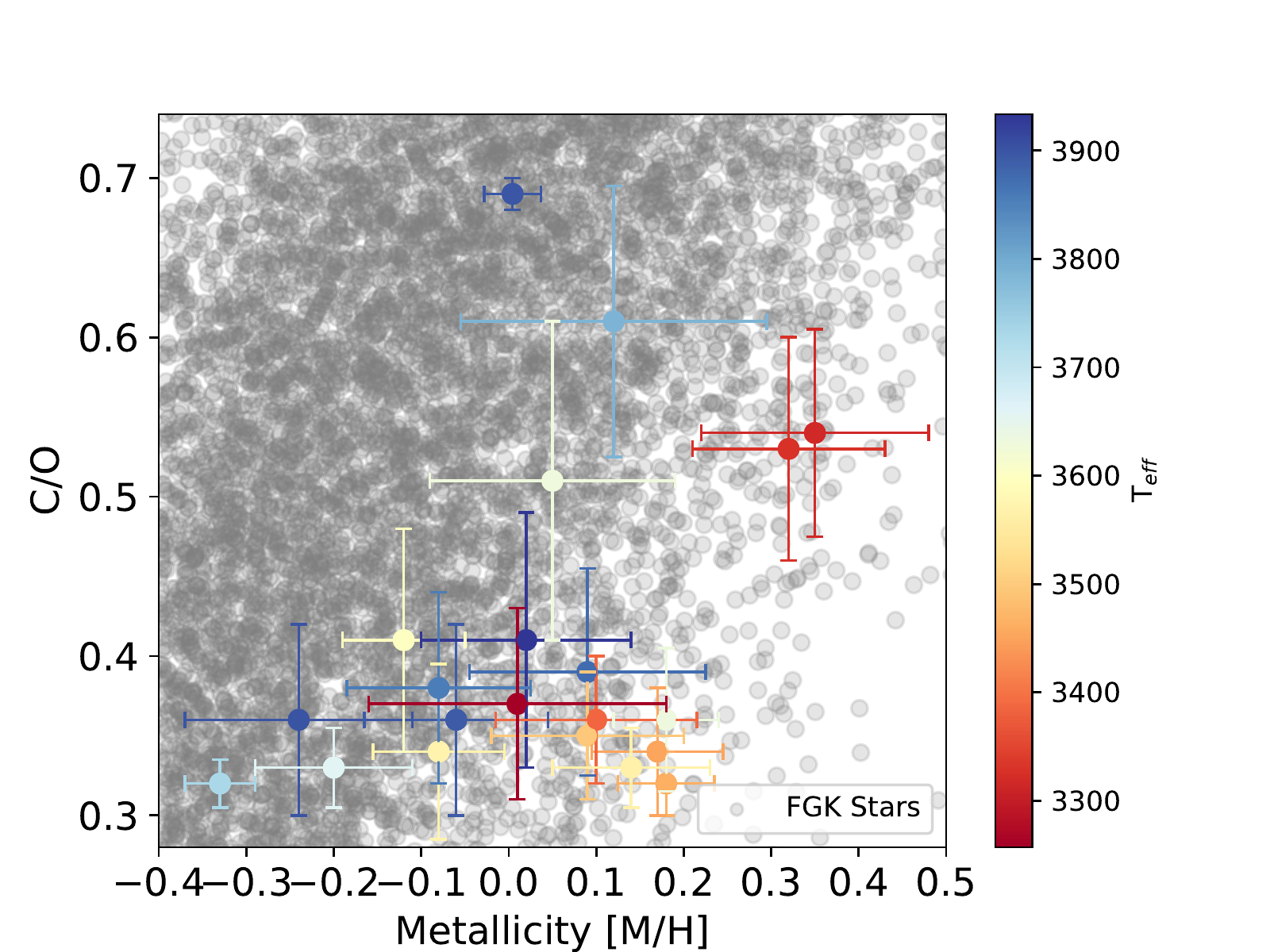}
    \centering
    \caption{Comparing our grid-model retrieved metallicity values against C/O for both sample groups (WBS + IS). We also show the derived effective temperatures for each target in the colorbar.In gray, we show local FGK population Metallicity and C/O estimates from \citep{hinkel2014}. Our derived values are consistent with the local population.}
    \label{metvsctoo}
  \end{figure}

Carbon-to-Oxygen Ratio is an extremely valuable composition dimension than can not only inform about the balance of carbon and oxygen species chemistry or refractory elements in atmospheres of cool stars but also about their own formation scenarios such as fragmentation \citep{offner2010formation} or core collapse \citep{gagne20182mass}. C/O ratios also shed light on the density, formation and history of migration of rocky planets or planetary mass companions harbored by M dwarfs \citep{gaidos2000cosmochemical}. Several retrieval techniques using low-resolution NIR spectra have acquired robust C/O constraints for cold brown dwarfs \citep{line2015uniform,line2017,zalesky2019,zalesky2022uniform,gonzales2020retrieval}; in addition to pEW \citep{nakajima2016carbon} and synthetic spectral model techniques \citep{gizis2016m,veyette2016physical} have been employed for M dwarfs. This work however; is a first in the simultaneous and systematic determination of [M/H] and C/O constraints of benchmark M dwarfs from low resolution spectra. In figure \ref{metvsctoo} we overlay our grid-model derived [M/H] and C/O values with local population FGK stars \citep{hinkel2014} (gray), showing overall consistency. A majority of our low mass star sample fall below the solar C/O value of 0.55. We also find that our average C/O constraints of $\sim$10$\%$ has not been previously achieved even for main-sequence G stars, specifically at low-resolution (R$\sim$120) spectra.

Deriving the C/O ratios of FGK stars have been challenging yet feasible due to limited blended lines in the spectra or accounting for Non-LTE corrections \citep{fortney2012carbon}. \citealt{mena2021} have computed C/O values for over 1000 FGK stars and found that derived C/O from high-resolution (R $>$ 100,000) spectra are extremely sensitive to the choice of oxygen line or stellar atmosphere models used to compute the ratios, leading to systematic shifts which need to be scaled with a solar reference value particularly for cool stars (K-type). In even cooler M dwarfs, the bulk of C goes into the CO opacity and the O into dominant metal oxide (TiO/VO) absorption features. Given that this falls typically in the optical/NIR region of the spectra that is more susceptible to model assumptions as seen from the above analysis, in future we plan to perform forward model Bayesian retrievals \citep{zalesky2022uniform,gonzales2020retrieval} to directly solve for the radiative-convective atmospheres of M dwarfs without having to rely on self-consistent grid-model assumptions.

\section{SUMMARY}\label{sec:summary}
 Numerous challenges plague the path of characterizing M dwarf atmospheres. In this work, we attempt to address on a fundamental level the need for improved stellar atmosphere models and the implications of working with outdated line lists; leading to problems that percolate into exoplanet characterization, galactic and chemical evolution and ultimately our understanding of the Universe. With such vast reaching applications of concern, here are our key takeaways:
 
 \begin{itemize}
     \item We present $\tt{SPHINX}$: a new 1-D radiative convective thermochemical equilibrium atmosphere model grid to characterize M dwarf atmospheres, permitting physical plausibility to solve for fundamental stellar parameters--T$_{eff}$, log$g$, [M/H], and R$_*$ using a Bayesian inference covariance matrix--error analysis method invoking Gaussian Processes ($Starfish$, \citealt{czekala_github,ian_czekala_2018_2221006}). We also include the most up-to-date compilation of molecular absorption cross-section data from EXOPLINES \citep{gharib2021exoplines}.
     \item We have benchmarked our stellar model grid with a grid-fitting routine described in section \ref{subsec:starfish} on low-resolution SpeX+SNIFS spectra of 10 widely-separated G+M binary targets and evaluated against empirically measured [M/H] \citep{mann2013metal}. Mainly, we compare how our model grid fares against PHOENIX-ACES \citep{husser2013} and find that not only are we consistent with the widely used model grid, we are also able to improve the [M/H] constraints relative to observations for 60$\%$ of our sample targets.
     \item We also benchmark our stellar atmosphere model grid by retrieving R$_*$ and T$_{eff}$ for low-resolution SpeX+SNIFS+STIS spectra of 10 targets with interferometrically measured angular diameters using the CHARA Array \citep{boyajian2012stellar}. Here we find that our retrieved parameters are consistent (with 95$\%$ and 99$\%$ confidence using 2-sample~K-S~test, see section \ref{results}) compared to observations as well as values derived using BT-Settl models \citep{mann2015constrain,allard2000spherically} with an average scatter of 4$\%$ and 5.2$\%$ respectively for T$_{eff}$ and R$_*$. We also find that our derived values are consistent within reasonable scatter compared to high-resolution spectral fitting as well as machine learning methods as surveyed and compiled in the form of literature median values by \citealt{passegger2022metallicities} (see Figure \ref{ALLRADCOMP}).
     \item For all 20 M dwarfs in this study, we are able to extend our grid to provide the stellar C/O ratio which is extremely valuable to assess the behavior of the carbon chemistry within the M dwarf as well as serve as valuable input parameters for planetary formation as well as disk evolution models \citep{fortney2012carbon}. We find that simultaneously constrained C/O values using our methodology for M dwarfs are overall consistent with neighboring FGK main sequence stars (see Figure \ref{metvsctoo}). In a future study, we also plan to expand our grid to provide N/C, $\alpha$/M ratios for M dwarfs.
     \item We also explore the possibility of stellar photospheric heterogeneity and inclusion of a simple linear combination parameterization for retrieving specifically the stellar spot temperature T$_{spot}$, the fractional coverage in area of the spots distributed on the photosphere f$_{spot}$ in addition to the stellar effective temperature T$_{eff}$. Assuming stellar photospheric heterogeneity as a hypothesis works well for certain targets and does not affect others, therefore a general consensus must be reached in order to understand the degree to which stellar heterogeneity would affect M dwarf abundances when focusing in the low-resolution (R$\sim$120) regime. We stick purely to low-resolution as we do not make assumptions for Non-LTE within our model grid and therefore present an elegant method of leveraging low-resolution spectra and acquiring constraints on bulk chemical properties such as C/O and [M/H].
     \item Lastly, we address the effect of general model assumptions that may also influence fundamental stellar properties that are retrieved such as the mixing length parameter $\alpha$. Here, we find that low-mass M dwarfs, particularly those that fall within the fully-convective regime are likely to be more sensitive to the assumed mixing-length parameter as opposed to the slightly hotter targets. However, assumed mixing length and therefore convection has a potential for degeneracies manifesting in the form of changing internal chemistry and ultimately the bulk metallicity and effective temperature, care must be taken especially when modeling low-mass stars. In a future study, we plan to perform free Bayesian retrievals (without self-consistent grid models and accompanying assumptions) on deriving appropriate atmospheric structures driven by low-resolution M dwarf data, and allowing for arbitrary species abundance determinations, which would truly stress-test model-assumptions such as these.
 \end{itemize}
 
 In conclusion, we present our new stellar atmosphere model grid and emphasize that low-resolution spectra show immense potential in acquiring robust constraints on fundamental properties of vital exoplanet hosting M dwarfs. Without invoking complicated Non-LTE models or 3-D to 1-D turbulent convection parameterizations, we are able to derive M dwarf effective temperatures (T$_{eff}$), surface gravity (log$g$), bulk chemical properties such as metallicity ([M/H]), carbon-to-oxygen ratio (C/O), and the stellar radius (R$_*$) all within reasonable scatter of values acquired from previous works (model-derived, empirical, and observational). We also gain insight into model-sensitive parameters that extend our caution into understanding and modeling cooler M dwarfs that fall within the sub-stellar boundary as well as implications into understanding brown dwarfs (late Ms and Ls) as we enter into the $JWST$ Era. In future works, we will extend this analysis to a larger homogeneous sample of M dwarf spectra in order to obtain broad population level trends in composition and fundamental parameters.

Supplementary figures are available at: \url{https://zenodo.org/record/6647451#.YrS6MJDMK3I}
 
\begin{table*}
\renewcommand{\thetable}{\arabic{table}}
\centering
\caption{Targets used for this analysis: 10 widely separated M+G Binaries, and 10 targets with interferometrically measured angular diameters. Values of fundamental stellar parameters retrieved from our grid-models.}\label{data}
\begin{tabular}{|c | c | c | c | c | c|}
\hline
Target  & T$_{eff}$ (K) & log$g$ (cm/s$^2$) & [M/H] & C/O & R$_*$ (R$_\odot$)\\
\hline
Wide Binary Sample (WBS) & & & \\
\hline
\decimals
\hline
NLTT 11270 & 3900.00$^{+56.28}_{-58.19}$ & 5.32$^{+0.15}_{-0.17}$ & -0.24$^{+0.14}_{-0.13}$ & 0.36$^{+0.06}_{-0.06}$ & 0.48$^{+0.01}_{-0.01}$\\
NLTT 11500 & 3599.93$^{+77.27}_{-61.72}$ & 4.92$^{+0.20}_{-0.21}$ & -0.12$^{+0.13}_{-0.13}$ & 0.41$^{+0.08}_{-0.06}$ & 0.36$^{+0.20}_{-0.21}$\\
NLTT 3725 & 3326.62$^{+82.61}_{-106.28}$ & 4.50$^{+0.12}_{-0.15}$ & 0.31$^{+0.10}_{-0.12}$ & 0.53$^{+0.07}_{-0.07}$ & 0.280$^{+0.002}_{-0.002}$ \\
NLTT 738  & 3630.88$^{+117.76}_{-73.35}$ &  4.65$^{+0.34}_{-0.26}$ & 0.05$^{+0.13}_{-0.15}$ & 0.51$^{+0.10}_{-0.10}$ & 0.28$^{+0.04}_{-0.04}$ \\
NLTT 8787 & 3874.58$^{+58.35}_{-94.11}$ & 5.28$^{+0.15}_{-0.24}$ & 0.09$^{+0.13}_{-0.14}$ & 0.39$^{+0.07}_{-0.06}$ & 0.34$^{+0.03}_{-0.03}$\\
Gl 100 & 3570.98$^{+77.93}_{-89.81}$ & 4.82$^{+0.21}_{-0.17}$ & -0.08$^{+0.07}_{-0.08}$ & 0.34$^{+0.08}_{-0.03}$ & 0.30$^{+0.03}_{-0.03}$ \\
Gl 105 & 3466.23$^{+56.70}_{-62.52}$ & 4.80$^{+0.12}_{-0.15}$ & 0.18$^{+0.06}_{-0.05}$ & 0.32$^{+0.03}_{-0.01}$ & 0.29$^{+0.12}_{-0.11}$\\
Gl 118.2 & 3316.50$^{+97.03}_{-106.3}$ & 4.77$^{+0.19}_{-0.16}$ &0.34$^{+0.09}_{-0.13}$ & 0.54$^{+0.06}_{-0.07}$ & 0.10$^{+0.05}_{-0.05}$\\
Gl 81.1 & 3996.01$^{+2.60}_{-4.45}$ & 5.24$^{+0.01}_{-0.01}$ & 0.00$^{+0.03}_{-0.03}$ & 0.69$^{+0.01}_{-0.01}$ & 0.71$^{+0.02}_{-0.02}$\\
NLTT 10349 & 3891.32$^{+65.43}_{-69.55}$ & 5.34$^{+0.11}_{-0.17}$ & -0.06$^{+0.10}_{-0.11}$ & 0.36$^{+0.08}_{-0.04}$ & 0.25$^{+0.01}_{-0.01}$\\

\hline
Interferometric Sample (IS) &  &  &\\
\hline
Gl 436 & 3633.71$^{+41.84}_{-47.42}$ & 5.19$^{+0.12}_{-0.13}$ & 0.18$^{+0.05}_{-0.07}$ & 0.36$^{+0.05}_{-0.04}$ & 0.44$^{+0.03}_{-0.03}$\\
Gl 880 & 3860.01$^{+27.21}_{-44.23}$ & 4.77$^{+0.33}_{-0.19}$ & -0.03$^{+0.13}_{-0.07}$ & 0.42$^{+0.10}_{-0.07}$ & 0.57$^{+0.01}_{-0.01}$\\
Gl 699 & 3450.15$^{+32.04}_{-62.51}$ & 4.97$^{+0.13}_{-0.16}$ & 0.17$^{+0.06}_{-0.09}$ & 034$^{+0.05}_{-0.03}$ & 0.19$^{+0.002}_{-0.002}$\\
Gl 526 & 3730.56$^{+18.40}_{-20.04}$ & 5.33$^{+0.06}_{-0.08}$ & -0.33$^{+0.04}_{-0.04}$ & 0.32$^{+0.02}_{-0.01}$ & 0.52$^{+0.01}_{-0.01}$\\
Gl 687 & 3497.45$^{+68.15}_{-71.32}$ & 5.0$^{+0.20}_{-0.25}$ & 0.09$^{+0.10}_{-0.12}$ & 0.35$^{+0.05}_{-0.03}$ & 0.41$^{+0.01}_{-0.01}$\\
Gl 887 & 3854.64$^{+31.69}_{-54.94}$ & 5.27$^{+0.17}_{-0.23}$ & -0.08$^{+0.11}_{-0.10}$ & 0.38$^{+0.07}_{-0.05}$ & 0.49$^{+0.001}_{-0.001}$\\
Gl 411 & 3656.81$^{+24.31}_{-31.92}$ & 5.26$^{+0.14}_{-0.15}$ & -0.2$^{+0.11}_{-0.06}$ & 0.33$^{+0.03}_{-0.02}$ & 0.38$^{+0.001}_{-0.001}$\\
Gl 581 & 3384.15$^{+67.20}_{-52.20}$ & 5.01$^{+0.21}_{-0.25}$ & 0.10$^{+0.10}_{-0.13}$ & 0.36$^{+0.05}_{-0.03}$ & 0.30$^{+0.001}_{-0.001}$\\
Gl 725A & 3563.91$^{+23.70}_{-36.89}$ & 5.27$^{+0.12}_{-0.20}$ & 0.14$^{+0.08}_{-0.10}$ & 0.33$^{+0.03}_{-0.02}$ & 0.35$^{+0.001}_{-0.001}$\\
Gl 725B & 3257.01$^{+44.80}_{-34.33}$ & 5.04$^{+0.30}_{-0.34}$ & 0.01$^{+0.15}_{-0.20}$ & 0.37$^{+0.07}_{-0.05}$ & 0.28$^{+0.001}_{-0.001}$\\
\hline
\hline
\end{tabular}
\label{tab:bestfit}
\begin{center}
\begin{minipage}{15.5cm}
\end{minipage}
\end{center}
\end{table*}

\acknowledgments

The authors would like to thank Andrew Mann for all the low resolution M dwarf spectra used in this work. ARI would also like to thank Ian Czekala, Miles Lucas, Michael Gully-Santiago, and Zhoujian (ZJ) Zhang for all their help and guidance in the working of $Starfish$. The authors would also like to credit NSF AAG Award, 2009592. ARI would like to acknowledge the NASA FINESST Grant 80NSSC21K1846.

This work has made use of and benefited from: Hypatia, Simbad, SpeX, Numpy, Scipy, Multinest

%

\vspace{5mm}
\facilities{SpeX SXD IRTF, SNIFS, HST-STIS}






\clearpage

\bibliography{biblio}
\bibliographystyle{aasjournal} 




\end{document}